\documentclass{emulateapj}
\usepackage{graphicx}
\begin{document}

\def\etal{et al.\ \rm}

\title{
Fast accretion of small planetesimals by protoplanetary cores.
}

\author{R. R. Rafikov}
\affil{IAS, Einstein Dr., Princeton, NJ 08540}
\email{rrr@ias.edu}


\begin{abstract}
We explore the dynamics of small planetesimals coexisting with 
massive protoplanetary cores in a gaseous nebula. Gas drag  
strongly affects the motion of small bodies leading to the 
decay of their eccentricities and inclinations, which are  
excited by the gravity of protoplanetary cores. Drag acting on
larger ($\gtrsim 1$ km), high velocity planetesimals 
causes a mere reduction of their 
average random velocity. By contrast, drag  
qualitatively changes the dynamics of smaller ($\lesssim 0.1-1$ km), 
low velocity objects: (1) small planetesimals sediment 
towards the midplane of the nebula 
forming vertically thin subdisk; (2) their 
random velocities rapidly decay between successive 
passages of the cores and, as a result, encounters 
with cores typically occur at the minimum relative 
velocity allowed by the shear in the disk. 
This leads to a drastic increase in the accretion rate of 
small planetesimals by the protoplanetary cores, allowing cores
to grow faster than expected in the simple oligarchic picture,  
provided that the population of small planetesimals contains 
more than roughly 1\% of the solid mass in the nebula. 
Fragmentation of larger planetesimals ($\gtrsim 1$ km) 
in energetic collisions triggered by the gravitational 
scattering by cores can easily channel this amount of 
material into small bodies on reasonable 
timescales ($< 1$ Myr in the outer Solar System), 
providing a means for the rapid growth (within several Myr
at 30 AU) of rather massive protoplanetary cores. Effects of 
inelastic collisions between planetesimals and presence
of multiple protoplanetary cores are discussed.
\end{abstract}



\section{Introduction.}\label{sect:intro}


Formation of terrestrial planets and solid cores 
of giant planets is thought to proceed via the 
gravity-assisted merging of a large number of 
planetesimals --- solid bodies with initial sizes of 
roughly several kilometers. Despite a considerable 
progress made in this field since 
the pioneering works of Safronov (1969), a number 
of important problems still remain unsolved.  
One of the most serious questions has to do with the
time needed for planets to complete their growth 
to present sizes. In the framework of conventional theory 
this time is rather long,
especially in the outer parts of the protoplanetary 
nebula ($\gtrsim 10^8-10^9$ yr), and it is  
likely that the gaseous component of the nebula 
dissipates much earlier (in $\lesssim 10^6-10^7$ 
yr). This would make it very hard for the 
giant planets in our Solar System to accrete their 
huge gaseous envelopes via core instability 
(Mizuno 1980) which is otherwise considered to be 
an attractive scenario.

Wetherill \& Stewart (1989) have identified 
a very rapid ``runaway'' regime of accretion 
of protoplanetary cores which at the time 
seemed like a solution of this problem. 
However, later on Ida \& Makino (1993) and 
Kokubo \& Ida (1996, 1998) have demonstrated that 
the runaway accretion would persist only through 
a rather limited interval of time and that 
the final growth of protoplanetary 
cores to isolation (which corresponds to roughly 
$10^{26}-10^{27}$ g at 1 AU) would proceed in a slow manner, 
making the formation of cores of giant planets 
rather problematic.  

These studies usually implied that the gaseous component 
of the nebula plays only a secondary role in the planet 
formation process. Planetesimals were 
typically assumed to be rather 
massive ($10^{23}-10^{24}$ g) bodies weakly 
affected by the gas
(Kokubo \& Ida 1998). This allows gravity to excite 
energetic random motions of planetesimals leading to 
diminishing the role of gravitational focusing and reduction 
of accretion efficiency. The purpose of this paper 
is to relax this assumption and to see what impact  
is incurred on the planet formation picture by allowing 
most of the planetesimals to be small ($\lesssim 10$ km) 
bodies immersed in a 
gaseous environment. Such planetesimals
would be appreciably affected by the gas drag and we will
demonstrate that this can bring qualitative changes 
to their dynamics in the 
vicinity of the protoplanetary cores and, consequently, 
to the behavior of the mass accretion rate of cores. 

Throughout this study we will use the following approximation 
to the structure of the Minimum Mass Solar Nebula (MMSN):
\begin{eqnarray}
&& \Sigma_g(a)\approx100\Sigma_p(a)\approx 3000~\mbox{g cm}^{-2}~a_{AU}^{-3/2},\\
&& c_s(a)\approx 1.2~\mbox{km s}^{-1}~a_{AU}^{-1/4},
\label{eq:MMSN} 
\end{eqnarray}
where $\Sigma_p, \Sigma_g$ are the particulate and gas surface 
densities correspondingly,  
$c_s$ is the gas sound speed, and $a_{AU}\equiv a/(1~\mbox{AU})$
is a distance from the Sun $a$ scaled by 1 AU. We will use the 
terms ``protoplanetary core'' and ``protoplanetary embryo''
interchangingly. Physical density of planetesimals $\rho_p$ is always
assumed to be $1$ g cm$^{-3}$.

The paper is organized as follows: after a discussion of
different gas drag regimes in \S \ref{sect:dissip} we
proceed to the description of planetesimal dynamics 
in the vicinity of protoplanetary cores in \S \ref{sect:scat}. 
The inclination of small planetesimals, a question 
very important for this study, is explored in 
\S \ref{subsect:inclination}.
The separation of different gas drag and dynamical 
regimes in different parts of the nebula is described 
in \S \ref{subsect:separation} and lower limits on 
the random velocities of planetesimals are obtained in 
\S \ref{sect:lower}. We dwell upon the role of inelastic collisions 
between planetesimals in \S \ref{subsect:inelastic}. 
The role of small planetesimals in the 
growth of protoplanetary cores is studied in 
\S \ref{sect:accretion} and some important
consequences for the planet 
formation picture are discussed in \S \ref{sect:discus}.


\section{Summary of different gas drag regimes.}\label{sect:dissip}


Drag force acting on a body moving in a gaseous medium 
depends on the relative velocity of the body with respect to 
the gas $v_r$ and on the ratio of its radius $r_p$
to the molecular mean free path $\lambda$ (Whipple 1972; 
Weidenschilling 1977). Whenever 
$r_p\lesssim\lambda$ the Epstein drag law applies:
\begin{eqnarray}
\frac{d{\bf v}_r}{dt}\simeq - \frac{\rho_g c_s}{\rho_p r_p}{\bf v}_r
\simeq-\Omega\frac{\Sigma_g}{\rho_p r_p}{\bf v}_r,
\label{eq:Epstein}
\end{eqnarray}
where $\Omega$ is the angular frequency of the disk, 
$\rho_g$ is the gas density, $r_p$ is the planetesimal size,
and $\rho_p$ is the physical density of planetesimals
(planetesimal random velocities are assumed to be subsonic).
For the adopted MMSN model (\ref{eq:MMSN}) 
we estimate 
\begin{eqnarray}
\lambda=(n_g\sigma_{H_2})^{-1}\simeq\frac{\mu c_s}{\Omega\Sigma_g
\sigma_{H_2}}
\simeq 1~\mbox{cm}~a_{AU}^{11/4}
\label{eq:lambda}
\end{eqnarray}
for H$_2$ collision cross section $\sigma_{H_2}\simeq 10^{-15}$
cm$^2$ ($\mu$ is the mean molecular weight). Here $n_g$ is a
number density of H$_2$ molecules.
It is important to notice that $\lambda$ increases 
very rapidly with the distance from the Sun, so that although 
only sub-cm particles can experience Epstein drag at 1 AU, 
at 30 AU this drag regime is valid even for rocks 100 m 
in size!

A large spherical body with $r_p\gtrsim\lambda$ experiences a  
deceleration of the form
\begin{eqnarray}
\frac{d{\bf v}_r}{dt}\simeq - C_D\frac{3}{4\pi} 
\frac{\rho_g}{\rho_p r_p} v_r{\bf v}_r.
\label{eq:non_Eps}
\end{eqnarray}
Drag coefficient $C_D$ is a function of the Reynolds number
$Re\equiv v_r r_p/\nu$, where $\nu$ is a kinematic viscosity. 
The viscosity we assume here is the molecular viscosity, i.e. 
$\nu\simeq \lambda c_s/3$; in the presence of anomalous 
sources of viscosity the expression for $\nu$ has to be 
correspondingly adjusted.

\begin{figure}
\plotone{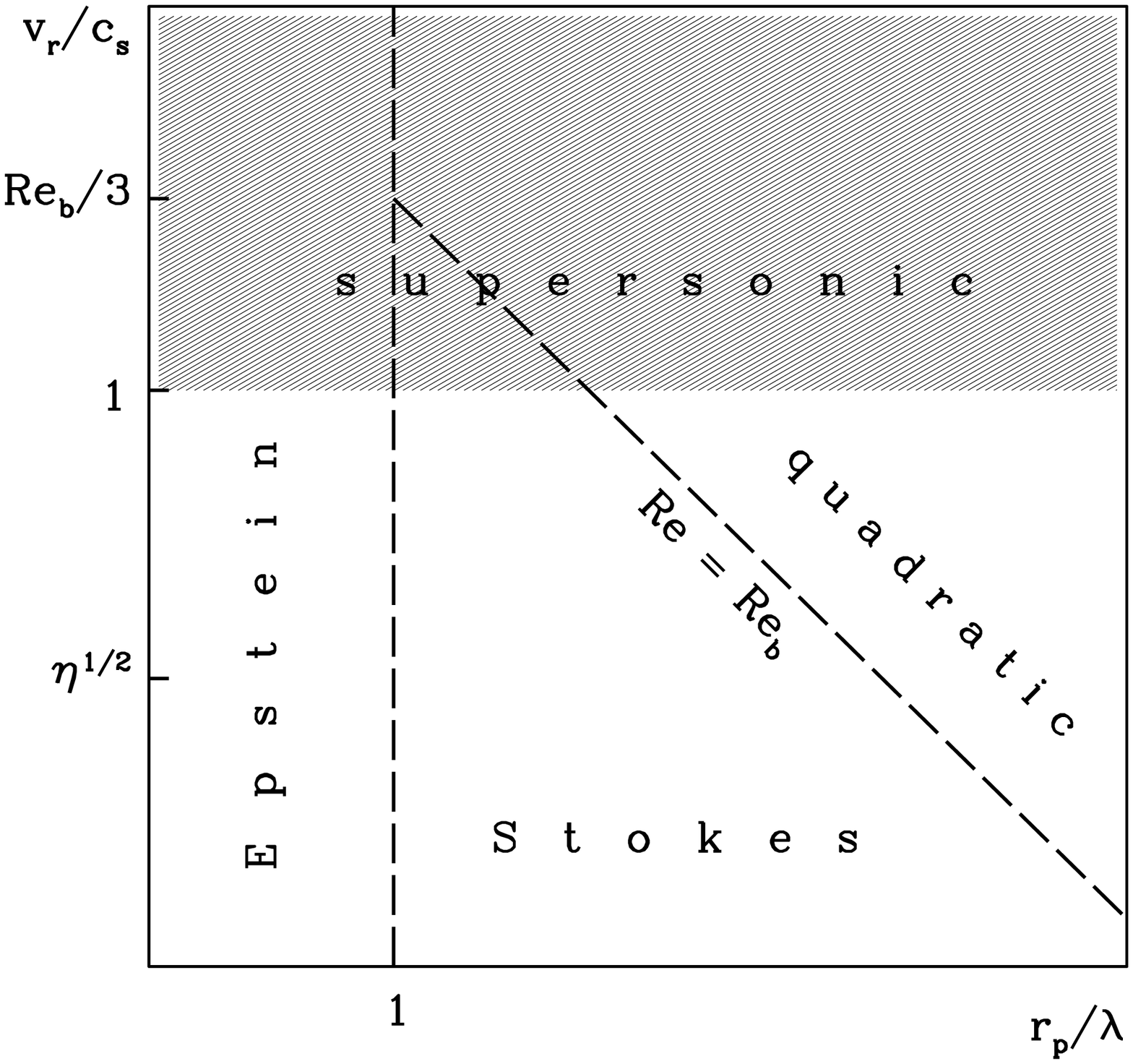}
\caption{
Separation of different gas drag regimes on $r_p/\lambda$ - 
$v_r/c_s$ plane. Dashed lines separate regions where different
gas drag laws (indicated on a plot) apply. Planetesimals 
in the shaded part of the plot move supersonically with respect
to the gas. Separation between the quadratic and Stokes regimes
($Re=Re_b$) is determined by $v_r/c_s=(Re_b/3)(r_p/\lambda)^{-1}$.
\label{fig:drag_regimes}}
\end{figure}

For $Re\lesssim 1$ gas drag is in 
the Stokes regime (Landau \& Lifshitz 1987), meaning that
$C_D=6\pi Re^{-1}$. For $Re\gg 1$ drag coefficient becomes
constant: $C_D\simeq 0.7$ (Weidenschilling 1977). We 
neglect the more complicated behavior of $C_D$ for
$1\lesssim Re\lesssim 10^2$ (Whipple 1972) 
and simply assume that $C_D$ switches from one
asymptotic behavior to the other at $Re_b\simeq 20$. Thus, we adopt 
\begin{eqnarray}
&& \frac{d{\bf v}_r}{dt}\simeq - \frac{3}{2}\Omega 
\frac{\Sigma_g}{\rho_p r_p}\frac{\lambda}{r_p}{\bf v}_r,~~~Re\lesssim Re_b,
\label{eq:Stokes}\\
&& \frac{d{\bf v}_r}{dt}\simeq - 0.2\Omega 
\frac{\Sigma_g}{\rho_p r_p}\frac{v_r}{c_s}{\bf v}_r,~~~Re\gtrsim Re_b.
\label{eq:large_vel}
\end{eqnarray}
Separation of different gas drag regimes 
as a function of $r_p$ and $v_r$ is schematically presented
in Figure \ref{fig:drag_regimes}.

Gas drag acting on a planetesimal moving on an elliptic and inclined 
orbit leads to the decay of planetesimal eccentricity $e$, 
inclination $i$, and semimajor axis $a$ (Adachi \etal 1976).
For the calculation of the decay rate of the orbital elements it
is important to take into account the sub-Keplerian angular 
velocity of the gas in the nebula caused by the 
radial pressure
support. This gives rise to additional azimuthal 
contribution $\Delta v_g$ to the relative gas-planetesimal velocity.
One can easily demonstrate that (Whipple 1971)
\begin{eqnarray}
&& \Delta v_g\approx \Omega a \left(\frac{c_s}{\Omega a}\right)^2\equiv 
\Omega a \eta
\simeq 50~\rm{m~s}^{-1},\\
&& \eta\equiv \left(\frac{c_s}{\Omega a}\right)^2\simeq
1.6\times 10^{-3}~a_{AU}^{1/2}.
\label{eq:delta_vg}
\end{eqnarray} 
For our adopted temperature profile $\Delta v_g$ is independent
of $a$. Whenever random
velocity of planetesimal exceeds $\Delta v_g$ 
this velocity offset provides only a 
small contribution to the relative
gas-planetesimal velocity. In the opposite case relative
velocity is dominated by $\Delta v_g$.

Using (\ref{eq:Epstein})-(\ref{eq:large_vel}) we can 
introduce a {\it damping time} $t_d$ --- typical time
needed to decelerate planetesimals by the gas drag if their 
initial velocity with respect to the gas is $\Delta v_g$. 
Using equations (\ref{eq:Epstein}), 
(\ref{eq:Stokes}), \& (\ref{eq:large_vel}) one obtains
\begin{eqnarray}
&& t_d\approx 5\Omega^{-1}\frac{\rho_p r_p}{\Sigma_g}
\frac{\Omega a}{c_s},
\label{eq:damp_quad}\\
&& t_d\approx \frac{2}{3}\Omega^{-1}\frac{\rho_p r_p}{\Sigma_g}
\frac{r_p}{\lambda},
\label{eq:damp_Stokes}\\
&& t_d\approx \Omega^{-1}\frac{\rho_p r_p}{\Sigma_g}
\label{eq:damp_Epstein}
\end{eqnarray}
for the quadratic, Stokes, and Epstein drag regimes respectively.

When $t_d\ll \Omega^{-1}$ planetesimal motion is tightly 
coupled to that of the gas; critical planetesimal size $r_{stop}$
at which the transition from almost Keplerian motion 
to the sub-Keplerian gas rotation occurs can be determined from
the condition $t_d\approx \Omega^{-1}$. Using 
(\ref{eq:damp_quad})-(\ref{eq:damp_Epstein}) we find that
\begin{eqnarray}
&& r_{stop}\approx 0.2\frac{\Sigma_g}{\rho_p}
\frac{c_s}{\Omega a}\approx 20~\mbox{cm}~a_{AU}^{-5/4},
\label{eq:stop_quad}\\
&& r_{stop}\approx \left(\frac{3}{2}\frac{\Sigma_g}{\rho_p}\lambda
\right)^{1/2}\approx 70~\mbox{cm}~a_{AU}^{5/8},
\label{eq:stop_Stokes}\\
&& r_{stop}\approx \frac{\Sigma_g}{\rho_p}
\approx 3\times 10^3~\mbox{cm}~a_{AU}^{-3/2},
\label{eq:stop_Epstein}
\end{eqnarray}
for the quadratic, Stokes, and Epstein drag regimes.

Calculation of the orbital evolution 
of planetesimals bigger than $r_{stop}$ due to gas drag   
was performed by Adachi \etal (1976; see also Tanaka \& Ida 1997)
taking into account the saturation of the relative gas-planetesimal 
velocity whenever the planetesimal random speed is $\lesssim\Delta v_g$. 
We quote their results for different drag regimes keeping only the most 
important contributions\footnote{We have set to unity all numerical 
factors which appear in the course of orbital averaging; however we
kept initial factors since some of them (e.g. in 
[\ref{eq:large_vel}]) are significantly different from unity.}:
\begin{eqnarray} 
&& \frac{1}{e^2}\frac{de^2}{dt}\approx \frac{2}{i^2}\frac{di^2}{dt}
\approx - 0.2~\Omega\frac{\Sigma_g}{\rho_p r_p}\frac{\Omega a}{c_s}
(e+i+\eta),
\label{eq:quadratic1}\\
&& \frac{1}{a}\frac{da}{dt}
\approx -0.2~\Omega\frac{\Sigma_g}{\rho_p r_p}\frac{\Omega a}{c_s}
\eta(e+i+\eta),
\label{eq:quadratic2}
\end{eqnarray}
for the quadratic dependence (\ref{eq:large_vel}) of gas 
drag on velocity. The decay rates represented by 
equations (\ref{eq:quadratic1}) 
\& (\ref{eq:quadratic2}) behave differently for 
$e,i\lesssim \eta$ and $e,i\gtrsim \eta$. In the first case the 
relative gas-planetesimal velocity is close to 
$\Delta v_g$; as a result, orbital elements exponentially decay 
with time. In the second case  $v_r$ is 
dominated by the planetesimal random motion and the 
decay time is inversely proportional to the random 
velocity $v\approx (e+i)\Omega a$.

For the linear dependence of gas drag on $v_r$ represented 
by the equations (\ref{eq:Epstein}) and (\ref{eq:Stokes}) 
one finds (Adachi \etal 1976)
\begin{eqnarray} 
&& \frac{1}{e^2}\frac{de^2}{dt}\approx \frac{2}{i^2}\frac{di^2}{dt}
\approx -\Omega\zeta\frac{\Sigma_g}{\rho_p r_p},
\label{eq:linear1}\\
&& \frac{1}{a}\frac{da}{dt}
\approx - \Omega\zeta\frac{\Sigma_g}{\rho_p r_p}\eta,
\label{eq:linear2}
\end{eqnarray}
where $\zeta=1$ for $r_p\lesssim \lambda$ (Epstein drag  
[\ref{eq:Epstein}]), and $\zeta=(3/2)\lambda/r_p$ for 
$\lambda\lesssim r_p$, $Re\lesssim Re_b$ (Stokes drag  
[\ref{eq:Stokes}]).


\section{Scattering by massive protoplanetary cores.}\label{sect:scat}


Let us consider the scattering of a planetesimal by a single 
protoplanetary core moving on a circular and uninclined orbit. 
The velocity perturbation which planetesimal receives 
in the course of gravitational interaction depends on the 
separation of the planetesimal guiding center
from the embryo's orbit and on the planetesimal approach 
velocity $v_a$. 

It is well known (H\'enon \& Petit 1986) that the Hill radius
$R_H$ of interacting bodies sets an important length scale for 
scattering in the disk. For two bodies 
with masses $m_1$ and $m_2$ Hill radius is defined as
\begin{eqnarray}
R_H\equiv a\left(\frac{m_1+m_2}{M_\odot}\right)^{1/3}.
\label{eq:Hill}
\end{eqnarray}
In the case of scattering of  planetesimals by a 
protoplanetary embryo having mass $M_e$ Hill radius is determined 
solely by $M_e$: $R_H=a(M_e/M_\odot)^{1/3}$. 

For our purposes the planetesimal approach velocity can be
represented as $v_a\approx \Omega(ea + ia+ R_H)$ and it is different 
from the horizontal random velocity $v_h\equiv e\Omega a$ 
and vertical random velocity $v_z\equiv i\Omega a$. Additional
contribution to $v_a$ in the form of the shear across the 
Hill radius $\Omega R_H$ (sometimes called the Hill velocity) 
appears because of the differential rotation of the disk. 
Gravitational focusing of planetesimals by the embryo is 
determined by $v_a$. Velocity $v_z$ determines 
the vertical thickness of planetesimal disk, thus it 
regulates the volume number density of planetesimals
$n_p$ for a given surface mass density
$\Sigma_p$: $n_p=\Omega \Sigma_p/(v_z m_p)$, where $m_p\equiv
(4\pi/3)\rho_pr_p^3$ is a 
planetesimal mass. Vertical velocity is 
much smaller than either $v_a$ or the total random velocity 
$v\approx v_h+v_z$ when $i\ll e$.

Whenever random velocity $v$ is less than 
$\Omega R_H$ --- the so-called
{\it shear-dominated}  regime (Ida 1990) --- close
approaches between gravitationally interacting bodies 
which might lead to their physical 
collision are only possible for the orbital separation 
$h\sim R_H$. For $h \gg R_H$ (distant encounters) 
orbital perturbations incurred in the course 
of scattering are small:
changes of the random velocity $\Delta v$ and of the 
orbital separation $\Delta h$ produced by a {\it single} 
scattering have magnitudes
\begin{eqnarray}
\Delta v\approx \Omega R_H\left(R_H/h\right)^2,~~~~
\Delta h\approx R_H\left(R_H/h\right)^2.
\label{eq:distant}
\end{eqnarray}
These increments can be positive or negative depending on the 
orbital phases of bodies and their random velocities before
the encounter. The averages of these quantities 
over epicyclic phases were calculated by Hasegawa \& Nakazawa 
(1990) and we quote here their result for the eccentricity
(it will be used in \S \ref{sect:lower}):
\begin{eqnarray}
\langle\Delta (e^2)\rangle\approx 5\left(R_H/a\right)^2
\left(\frac{R_H}{h}\right)^4.
\label{eq:distant_av}
\end{eqnarray}

In the case $h\sim R_H$, when interacting bodies enter their 
mutual Hill sphere, strong scattering takes place:
\begin{eqnarray}
\Delta v\approx \Omega R_H,~~~~\Delta (h^2)\approx R_H^2, 
\label{eq:close}
\end{eqnarray} 
with both increments being positive --- orbits are 
repelled and random velocity  $\sim\Omega R_H$
is excited. At the same time, it is clear from purely geometrical 
considerations that for $v\ll \Omega R_H$ the change of the vertical 
velocity is proportional to 
the vertical projection of the random velocity perturbation, 
and is much smaller than the change in the horizontal velocity:
\begin{eqnarray}
\Delta v_z\sim \Delta v\frac{v_z}{v_a}\approx v_z,~~\mbox{or}~~
\Delta i\sim i.
\label{eq:close_incl}
\end{eqnarray}
Since the total velocity increment $\Delta v$ is of the order or bigger 
than the initial random velocity of planetesimals, scattering
for $h\sim R_H$ in the shear-dominated regime has a {\it discrete} 
character.

Another possible dynamical state of planetesimals  
is the {\it dispersion-dominated}
regime which takes place whenever $v\gtrsim \Omega R_H$. In this case 
the epicyclic excursions of planetesimals allow collisions between
the bodies with orbits separated by $h<v/\Omega$; more distant bodies 
are again subject to only weak scattering. 
Moreover, even particles
having close approaches experience {\it on average} 
only small velocity 
perturbation compared to the pre-encounter velocity $v_a$. This 
allows one to treat the scattering in the dispersion-dominated 
regime as a {\it continuous} process, unlike the discrete 
scattering in the
shear-dominated regime. Another important feature of the 
dispersion-dominated case is that $v_z\sim v_h$, or $i\sim e$,
as a result of roughly three-dimensional nature of 
scattering in this  regime which tends to isotropize 
highly anisotropic velocity distributions
(Ida, Kokubo, \& Makino 1993; Rafikov 2003c).
Thus, it is enough to follow the evolution of only 
one component of planetesimal velocity (e.g. $e$) 
in the dispersion-dominated regime. 


\subsection{Quadratic drag.}
\label{subsect:high_vel}


Let's now proceed to considering planetesimal dynamics for  
different gas drag regimes. First, we explore 
the case of quadratic drag represented by 
equations (\ref{eq:large_vel}), (\ref{eq:quadratic1}),  and 
(\ref{eq:quadratic2}) which is valid for $r_p\gtrsim \lambda,  
Re\gtrsim Re_b$. This situation would most 
likely be realized in the inner parts of the protoplanetary 
nebula since $\lambda$ is rather
small there making Epstein regime irrelevant and bringing 
Reynolds number to a high value. 

We first assume that scattering occurs in the 
dispersion-dominated regime and try to figure out under 
which conditions a 
steady state can be realized in this case. 
Gravitational 
scattering by the embryo increases planetesimal random 
energy at a rate\footnote{In the dispersion-dominated regime 
$v_a\approx v$.} (Ida \& Makino 1993; Rafikov 2003b)
\begin{eqnarray} 
\frac{dv^2}{dt}\approx \Omega(\Omega R_H)^2\frac{R_H}{a}
\left(\frac{\Omega R_H}{v}\right)^3\ln\Lambda,
\label{eq:scat_rate}
\end{eqnarray}
where $\ln\Lambda\sim 1$ is a Coulomb logarithm and 
$\Lambda\approx \left(v/\Omega R_H\right)^3$ 
for $e\sim i \gtrsim R_H/a$ (Stewart \& Ida 2000).
Because of the very weak dependence of $\ln\Lambda$ on 
velocity we will set it to unity in our further discussion.

Balancing this growth rate by the damping due to the gas drag 
(equation [\ref{eq:quadratic1}]) we find that
\begin{eqnarray}
&& v\approx \Omega R_H\left(5\frac{c_s}{\Omega a}\frac{\rho_p r_p}{\Sigma_g}
\right)^{1/6}\nonumber \\
&& \approx 1.4~\Omega R_H\left(\frac{r_p}{1~\mbox{km}}
\right)^{1/6} a_{AU}^{7/24},~~~~\mbox{for}~~v\gtrsim \Delta v_g,
\label{eq:bigger_offset}\\
&& v\approx \Omega R_H\left(5\frac{c_s}{\Omega a}\frac{\rho_p r_p}{\Sigma_g}
\frac{R_H}{\eta a}\right)^{1/5}\nonumber \\
&& \approx 
1.5~\Omega R_H\left(\frac{r_p}{1~\mbox{km}}
\right)^{1/5}\left(\frac{M_e}{10^{25}~\mbox{g}}
\right)^{1/15} a_{AU}^{1/4},\nonumber\\
&& ~~~~~~~~~~~~~~~~~~~~~~~~~~~~~~~~~~~~~~~~~\mbox{for}~~v\lesssim \Delta v_g.
\label{eq:smaller_offset}
\end{eqnarray}

For further convenience we normalize the embryo's mass $M_e$ to a fiducial
mass $M_f$ and planetesimal radius to a fiducial radius $r_f$ which 
are defined as
\begin{eqnarray}
&& M_f\equiv M_\odot \eta^3\simeq 8\times 10^{24}~\mbox{g}~a_{AU}^{3/2},
\label{eq:M_f}\\
&& r_f\equiv 0.2 \frac{\Sigma_g}{\rho_p}\frac{\Omega a}{c_s}=0.2 a\frac{\rho_g}{\rho_p}
\simeq 150~\mbox{m}~a_{AU}^{-7/4}.
\label{eq:r_f}
\end{eqnarray}
These two parameters uniquely characterize the dynamics
of the embryo-planetesimal scattering when the gas drag 
dependence on the 
velocity is represented by equation (\ref{eq:large_vel}).  
Using this notation we conclude from (\ref{eq:bigger_offset})
that $v\gtrsim \Omega R_H$ and $v\gtrsim \Delta v_g$ if
\begin{eqnarray}
&& r_p/r_f\gtrsim 1,
\label{eq:dd_cond}\\
&& M_e/M_f\gtrsim \left(r_p/r_f\right)^{-1/2}.
\label{eq:high_vel_cond}
\end{eqnarray}
From (\ref{eq:smaller_offset}) we find that 
$v\gtrsim \Omega R_H$ and $v\lesssim \Delta v_g$ if
\begin{eqnarray}
&& M_e/M_f\gtrsim \left(r_p/r_f\right)^{-3},
\label{eq:dd_cond1}\\
&& M_e/M_f\lesssim \left(r_p/r_f\right)^{-1/2}.
\label{eq:low_vel_cond}
\end{eqnarray}
Thus, conditions (\ref{eq:dd_cond}) and (\ref{eq:dd_cond1}) 
determine the region in the parameter space $M_e-r_p$ in 
which planetesimals are dispersion-dominated 
with respect to embryo and maintain a steady-state
velocity dispersion, see Figure \ref{fig:earth}.

Whenever both (\ref{eq:dd_cond}) and 
(\ref{eq:dd_cond1}) are not fulfilled, 
scattering of planetesimals proceeds in the 
shear-dominated regime. As we have previously mentioned, this type
of interaction has a discrete nature --- gas drag causes  significant
evolution of planetesimal velocity between close approaches 
to the embryo. We consider a planetesimal initially
separated in semimajor axis from the embryo by $h\sim R_H$ with
initial random velocity $v\lesssim \Omega R_H$. Immediately 
after scattering $v$ increases 
to $\sim \Omega R_H$, and the post-scattering value of
planetesimal eccentricity is $e_0\sim R_H/a=\eta(M_e/M_f)^{1/3}$.

We will see in \S \ref{subsect:inclination} that the 
inclination of planetesimals interacting
with the embryo in the shear-dominated regime is typically much 
smaller than their eccentricity. Then we can easily solve equation 
(\ref{eq:quadratic1}) with initial condition $e(0)=e_0$ 
to obtain a general solution in the form 
\begin{eqnarray}
e(t)\approx \eta\left[\left(1+\eta/e_0\right)\exp\left(
\Omega t\eta\frac{r_f}{r_p}\right)-1\right]^{-1}.
\label{eq:ecc_ev}
\end{eqnarray}

For low-mass embryos, $M_e\lesssim M_f$, planetesimal velocity 
after shear-dominated scattering is always less than $\Delta v_g$, 
i.e $e_0\lesssim \eta$;
as a result, we find from (\ref{eq:ecc_ev}) that $v$ damps exponentially:
\begin{eqnarray}
e(t)\approx e_0\exp\left(-\Omega t\eta\frac{r_f}{r_p}\right)
\approx\eta\left(\frac{M_e}{M_f}\right)^{1/3}e^{-t/t_d}.
\label{eq:ecc_ev_exp}
\end{eqnarray}
Thus, eccentricity decays on a typical 
timescale $t_d$, see (\ref{eq:damp_quad}) \& (\ref{eq:r_f}).
   
For high-mass embryos, $M_e\gtrsim M_f$, post-encounter velocity
is above $\Delta v_g$ ($e_0\gtrsim \eta$) and damping by 
the gas drag is very efficient:
$dv^2/dt\propto -v^3$, see equation (\ref{eq:large_vel}). One can
find from (\ref{eq:ecc_ev}) that in this case
\begin{eqnarray}
e(t)\approx \frac{\eta}{(\eta/e_0)+\Omega t\eta(r_f/r_p)}=
\frac{\eta}{(\eta/e_0)+(t/t_d)},
\label{eq:ecc_ev_pow}
\end{eqnarray}
for $t\lesssim t_d$. One finds that $e$ drops to $\eta$ 
after $t\approx t_d$ independent of $e_0$. 
Beyond that point relative gas-planetesimal
velocity is set by $\Delta v_g$, thus for $t\gtrsim 
t_d$ eccentricity decays exponentially on a timescale 
$t_d$, analogous to (\ref{eq:ecc_ev_exp}). It also follows from 
(\ref{eq:ecc_ev_pow}) that significant
reduction of eccentricity after scattering by the high mass 
embryo occurs already on a timescale 
$t_d (\eta/e_0)
\approx \Omega^{-1}(r_p/r_f)(a/R_H)\ll t_d$.

Planetesimals radially separated  from the embryo by
$\sim R_H$  pass the embryo's Hill sphere
(and experience strong scattering) roughly every synodic period 
$t_{syn}$ corresponding to the radial separation of $R_H$: 
\begin{eqnarray}
t_{syn}\approx \Omega^{-1}(a/R_H).
\label{eq:synodic}
\end{eqnarray}
Requiring the damping time $t_d$ to be shorter than $t_{syn}$ 
(which is necessary for planetesimals to stay in the shear-dominated regime) 
for embryos less massive than $M_f$ is equivalent to demanding that
\begin{eqnarray}
\eta^{-1}(r_p/r_f)\lesssim a/R_H~~~\mbox{or}~~~
M_e/M_f\lesssim (r_p/r_f)^{-3}. 
\label{eq:sd_cond}
\end{eqnarray}
Around more massive embryos with $M_e\gtrsim M_f$ planetesimal 
eccentricity is 
strongly damped between consecutive encounters with the 
embryo below its initial value 
 if $t_d (\eta/e_0)\lesssim t_{syn}$ or if
\begin{eqnarray}
(r_p/r_f)(a/R_H)\lesssim a/R_H~~~\mbox{or}~~~
r_p/r_f\lesssim 1. 
\label{eq:sd_cond1}
\end{eqnarray}
At the same time, eccentricity would not drop below 
$\eta$ prior to the next encounter if 
(\ref{eq:sd_cond}) is not fulfilled simultaneously with
(\ref{eq:sd_cond1}). 

Comparing (\ref{eq:sd_cond}) and (\ref{eq:sd_cond1}) with 
the dispersion-dominated conditions (\ref{eq:dd_cond}) and
(\ref{eq:dd_cond1}) we see that depending on 
$M_e/M_f$ and $r_p/r_f$ there can only be two
possible states in the system: (1) either planetesimals are
scattered in a smooth, continuous fashion in the 
dispersion-dominated regime, with gas drag not capable of 
damping their eccentricities significantly 
between the consecutive approaches 
to the embryo, or (2) they are strongly scattered by the 
embryo in the shear-dominated regime at 
each approach and gas appreciably reduces their random 
velocities before the next encounter takes place.

The discrete nature of planetesimal scattering 
in the shear-dominated regime is very important for determining
the approach velocity of planetesimals to the embryo. For example,
if we were to assume that scattering in the shear-dominated 
regime is continuous (like in the dispersion-dominated case) 
the average rate of eccentricity growth would have had the form 
$de/dt\sim (R_H/a)/t_{syn}$, since embryo increases planetesimal
eccentricity by $\sim R_H/a$ every synodic period. Balancing this
by the gas drag in the form (\ref{eq:large_vel}), one would find
the average value of eccentricity to be $\sim (R_H/a)
(t_d/t_{syn})$.
It would however be a grave mistake to assume that this is the
eccentricity with which planetesimal approaches the embryo. 
Indeed, it follows from (\ref{eq:ecc_ev_exp}) that the 
planetesimal eccentricity right before the encounter 
with the embryo is $\sim (R_H/a)\exp(-t_{syn}/t_d)$ for
$M_e\lesssim M_f$, which is exponentially smaller than the 
average value of $e$! Thus, proper
treatment of the shear-dominated regime 
taking the discrete nature of scattering 
into account is crucial for
figuring out the initial conditions of the interaction 
process. This will have important ramifications for 
the question of accretion of these planetesimals as 
we demonstrate in \S \ref{sect:accretion}.


\subsection{Stokes drag.}
\label{subsect:Stokes}


Planetesimals interact with the gas in the Stokes regime
when $r_p\gtrsim \lambda$ and $Re\lesssim Re_b$. The last 
condition depends not only on the particle size $r_p$ but
also on its velocity. We introduce another
fiducial size, $r_S$, which is defined as 
a planetesimal size for which $Re=Re_b$ 
at $v=\Delta v_g$:
\begin{eqnarray}
r_S\equiv \lambda\frac{Re_b}{3}\frac{\Omega a}{c_s}\approx
2~\mbox{m}~a_{AU}^{5/2}.
\label{eq:r_S}
\end{eqnarray}
Planetesimals with $r_p\lesssim r_S$ always experience 
gas drag in the Stokes regime for $M_p\lesssim M_f$ (because
post-scattering planetesimal velocity is 
$\lesssim \Delta v_g$, $v_r\approx \Delta v_g$,
and $Re\lesssim Re_b$).

Embryos more massive than $M_f$ endow shear-dominated 
planetesimals with velocity $\Omega R_H>\Delta v_g$ at each
scattering episode. But even 
then planetesimals with sizes $r_p\lesssim r_S$ satisfying
condition
\begin{eqnarray}
M_e/M_f\lesssim (r_p/r_S)^{-3}=\left(\frac{r_S}{r_f}\right)^3
(r_p/r_f)^{-3}
\label{eq:always_Stokes}
\end{eqnarray}
experience {\it only} the Stokes drag, 
i.e. their maximum 
velocity and physical size are never large enough 
for their Reynolds number to exceed 
$Re_b$. Planetesimals scattered by more massive 
embryos would experience quadratic drag right 
after the encounter (even if 
only temporarily) before switching to the 
Stokes regime. From equation
(\ref{eq:linear1}) one can easily see that damping
between encounters is purely exponential for the Stokes drag
independent of the planetesimal velocity (similar to
the quadratic drag for $e\lesssim \eta$, see \S 
\ref{subsect:high_vel}). 

Let us now turn to the dispersion-dominated 
regime. Balancing heating
rate (\ref{eq:scat_rate}) by the damping rate 
(\ref{eq:linear1}) with $\zeta=(3/2)\lambda/r_p$ we find that
\begin{eqnarray}
&& v\approx \Omega R_H\left[\frac{2}{3}\left(\frac{M_e}{M_f}\right)^{1/3}
\left(\frac{r_p}{r_f}
\right)^2\frac{r_f}{r_S}\right]^{1/5}\nonumber\\
&& \approx
5~\Omega R_H\left(\frac{r_p}{1~\mbox{km}}\right)^{2/5}
\left(\frac{M_e}{10^{25}~\mbox{g}}\right)^{1/15}a_{AU}^{-1/4}.
\label{eq:vel_Stokes}
\end{eqnarray}
Consequently, planetesimals are in the dispersion-dominated regime 
with respect to the embryo if
\begin{eqnarray}
M_e/M_f\gtrsim \left(\frac{r_S}{r_f}\right)^3
(r_p/r_f)^{-6}.
\label{eq:Stokes_restr}
\end{eqnarray}
Smaller planetesimals are in the shear-dominated regime and 
experience discrete scattering by the embryos. 
From equation (\ref{eq:vel_Stokes}) we also find that 
$v\lesssim \Delta v_g$ whenever
\begin{eqnarray}
M_e/M_f\lesssim \left(\frac{r_S}{r_f}\right)^{1/2}(r_p/r_f)^{-1},
\label{eq:Stokes_vg}
\end{eqnarray}
and that $Re< Re_b$ (and drag is in the Stokes regime) if
\begin{eqnarray}
M_e/M_f\lesssim \left(\frac{r_S}{r_f}\right)^{3}(r_p/r_f)^{-7/2}
\label{eq:Stokes_bnd}
\end{eqnarray}
(see Figure \ref{fig:jupiter}).
The last equation is a dispersion-dominated analog of the condition
(\ref{eq:always_Stokes}).


\subsection{Epstein drag.}
\label{subsect:Epstein}


Smallest planetesimals with $r_p\lesssim \lambda$,
are coupled to gas via the Epstein drag. Performing analysis 
analogous to that of \S \ref{subsect:Stokes} one finds that
in the dispersion-dominated regime scattering by the
embryo maintains planetesimal random velocity at the level of
\begin{eqnarray}
&& v\approx \Omega R_H\left[\left(\frac{M_e}{M_f}\right)^{1/3}
\frac{\lambda}{r_S}\frac{r_p}{r_f}\right]^{1/5}\nonumber\\
&& \approx
1.4~\Omega R_H
\left(\frac{r_p}{1~\mbox{km}}\right)^{1/5}
\left(\frac{M_e}{10^{25}~\mbox{g}}\right)^{1/15}
\left(\frac{a_{AU}}{30}\right)^{3/10}.
\label{eq:vel_Epstein}
\end{eqnarray}
Planetesimals can only be 
in the dispersion-dominated regime with respect to the embryo if 
\begin{eqnarray}
M_e/M_f\gtrsim \left(\frac{r_S}{\lambda}\right)^3
(r_p/r_f)^{-3}
\label{eq:Epstein_restr}
\end{eqnarray}
(see Figure \ref{fig:neptune}).
In this dynamical regime planetesimal velocity is below
$\Delta v_g$ only if 
\begin{eqnarray}
M_e/M_f\lesssim \left(\frac{r_S}{\lambda}\right)^{1/2}
(r_p/r_f)^{-1/2}.
\label{eq:Epstein_vg}
\end{eqnarray}

Planetesimals too small to satisfy (\ref{eq:Epstein_restr}) are 
in the shear-dominated regime and experience strong scattering 
by the embryo every synodic period, with their orbital elements 
exponentially decaying between encounters
(analogous to the behavior in the case of Stokes 
drag, see \S \ref{subsect:Stokes}). 


\subsection{Inclination evolution.}
\label{subsect:inclination}


It is easy to see from equations (\ref{eq:quadratic1}) 
and (\ref{eq:linear1}) that
as long as gas drag is the only force affecting 
planetesimals after their encounter with embryo, planetesimal 
inclination decays according to
\begin{eqnarray}
i(t)\approx i_0\sqrt{e(t)/e_0},
\label{eq:incl_decay}
\end{eqnarray}
with $e_0, i_0$ being the post-encounter values of eccentricity
and inclination and $e(t)$ given by (\ref{eq:ecc_ev})
for quadratic gas drag. In the shear-dominated regime 
one might be tempted to think on the basis of 
(\ref{eq:incl_decay}) that $i\gg e$ 
long after the encounter took place (when $e\ll e_0$). 
This argument, however, assumes that
$i_0\sim e_0$ and we now show that this is not the case.

\begin{figure}
\plotone{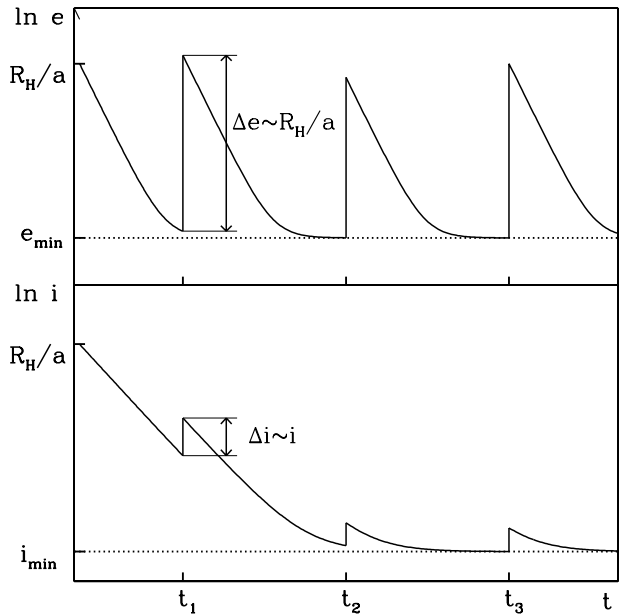}
\caption{
Sketch of the planetesimal eccentricity ({\it top}) and 
inclination ({\it bottom}) evolution due to the 
scattering by the embryo. Spikes correspond to close 
approaches to the embryo (which happen at different intervals
because of the changing semimajor axis separation). We assume that 
there is a weak continuous source of planetesimal excitation
keeping random velocities above some minimum level ($e_{min}$
and $i_{min}$). In reality behaviors of $e$ and $i$ are more 
erratic because of the scattering by distant embryos (see \S 
\ref{sect:lower}).  
\label{fig:inclination}}
\end{figure}

Indeed, even if $i_0\sim R_H/a$, inclination 
is exponentially small right before the next encounter 
(see [\ref{eq:ecc_ev_exp}] and [\ref{eq:incl_decay}])
since the damping time is shorter than the synodic period 
in the shear-dominated regime. After this second encounter,
eccentricity increases to $\sim R_H/a$, while the inclination does 
not go back to $R_H/a$ but, 
according to (\ref{eq:close_incl}), remains small, see Figure 
\ref{fig:inclination}. Subsequent action of gas drag prior to  
the next passage of the core further reduces planetesimal 
inclination, and so on. As a result, we arrive at 
a very interesting conclusion: as long as planetesimals 
are in the shear-dominated regime their inclinations keeps
decaying. Thus, all shear-dominated planetesimals 
``rain out'' towards the disk midplane and collapse into  
geometrically thin layer. This emphasizes the importance of 
determining which part of the planetesimal population is 
shear-dominated with respect to the embryo. 

Thickness of this layer would be 
zero if embryo were on purely uninclined 
orbit and gas drag were the only process affecting planetesimal
dynamics between encounters. In reality, there 
are additional stirring agents which
would keep the thickness of this subdisk finite (although 
still very small) and we consider them in \S \ref{sect:lower}
\& \S \ref{subsect:inelastic}.


\subsection{Separation of different regimes.}
\label{subsect:separation}


We now summarize what we have learned in \S 
\ref{subsect:high_vel}-\ref{subsect:Epstein} about 
planetesimal dynamics in the 
proto-Solar nebula.

\begin{figure}
\plotone{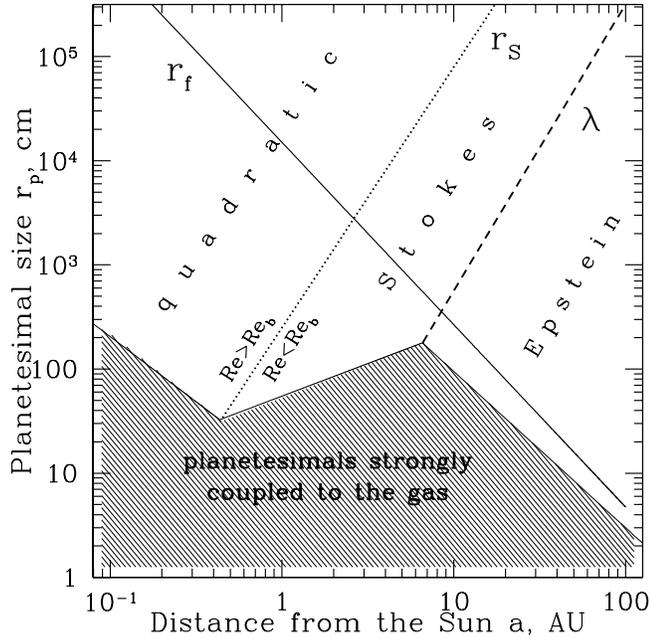}
\caption{
Behavior of the length scales $r_f$, $r_S$, and $\lambda$ important 
for the planetesimal interaction with the gas (with different regimes 
marked on the plot), as a function of distance $a$ from the Sun. 
Planetesimals with sizes in the shaded region 
have stopping time shorter than local $\Omega^{-1}$ and are thus
moving together with the gas. Complex shape of the boundary of 
this region is due to the different gas drag regimes.
\label{fig:length_scales}}
\end{figure}

First of all, it is clear from our previous discussion  
that the separation of different gas drag regimes sensitively
depends on the relative values of fiducial planetesimal
sizes $\lambda$, $r_f$, and $r_S$ given by the equations 
(\ref{eq:lambda}), (\ref{eq:r_f}), and (\ref{eq:r_S}). In Figure 
\ref{fig:length_scales} we display the scaling of these 
sizes with the distance from the Sun. The boundary 
between the quadratic and Stokes drag regimes is calculated assuming
$v\approx \Delta v_g$ and coincides with $r_S(a)$. 
We also display the curve $r_{stop}(a)$ using equations 
(\ref{eq:stop_quad})-(\ref{eq:stop_Epstein}) --- 
planetesimals with $r_p$ below this curve (shaded region) are tightly
coupled to the gas and their dynamics cannot be described by equations
(\ref{eq:quadratic1})-(\ref{eq:linear2}).

From Figure \ref{fig:length_scales} one can see that depending on the
location in the nebula there could be three possible situations:
(a) when $\lambda\lesssim r_S\lesssim r_f$, which holds 
for $a\lesssim 3$ AU in the MMSN, (b) when 
$\lambda\lesssim r_f\lesssim r_S$, which takes place for $3$ AU 
$\lesssim a\lesssim 9$ AU, and (c) when 
$r_f\lesssim\lambda\lesssim r_S$, which is valid for 
$a\gtrsim 9$ AU.

\begin{figure}
\plotone{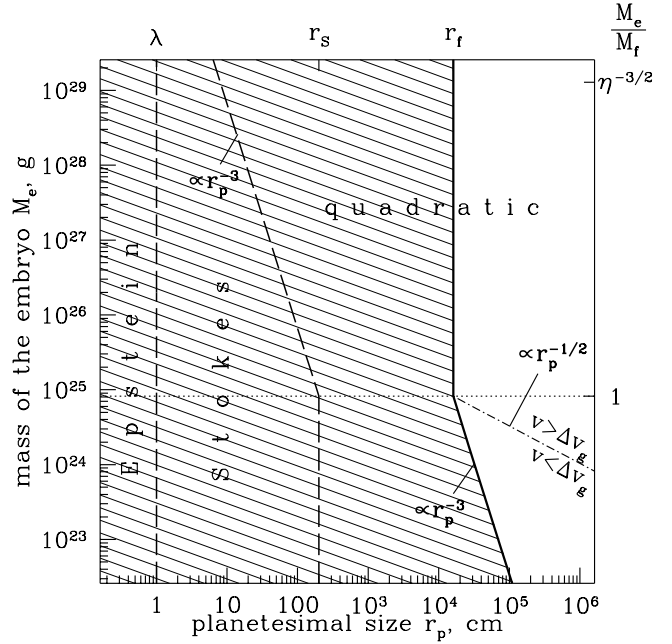}
\caption{
Separation of planetesimals into the shear- and dispersion-dominated 
ones with
respect to the embryo in the phase space of the 
embryo's mass $M_e$ and
planetesimals size $r_p$. This particular separation and numerical 
values indicated in the Figure pertain 
$a=1$ AU (terrestrial planet region), where 
$\lambda\lesssim r_S\lesssim r_f$. Thick solid line separates 
two dynamical regimes, with shear-dominated one being hatched. 
Thick dashed curves separate different regimes of planetesimal 
interaction with the gas (explicitly indicated on the plot). 
See text for more details. 
\label{fig:earth}}
\end{figure}

The first case pertains to the region of terrestrial planets. 
We display possible regimes of planetesimal interaction 
with gas in this part of the nebula in Figure \ref{fig:earth}. 
They are classified according to the values of planetesimal
size $r_p$ and embryo mass $M_e$. Thick solid line separates 
planetesimals which interact with the embryo in the 
shear-dominated regime and experience discrete scattering 
(hatched region to the left of the curve)
from those which undergo the  
dispersion-dominated scattering (unhatched region). In the terrestrial 
region, as we see from Figure \ref{fig:earth}, this curve
is set by equations (\ref{eq:dd_cond}) and (\ref{eq:dd_cond1}). 
Dashed curves denote the 
boundaries of different gas drag regimes: Epstein drag operates 
when $r_p\lesssim \lambda$; Stokes drag operates 
whenever $Re\lesssim Re_b$ --- in the shear-dominated regime 
this implies $r_p\lesssim r_S$ for embryos that cannot kick 
planetesimals by more than $\Delta v_g$ ($M_e\lesssim M_f$, 
see \S \ref{subsect:Stokes}), and restriction (\ref{eq:always_Stokes}) 
for embryos that can\footnote{In fact, even near embryos 
more massive than $M_e$ determined from (\ref{eq:always_Stokes}) 
planetesimals can spend some time in the Stokes regime: 
although they are in the quadratic drag regime right after the 
passage of the embryo, their velocity rapidly decreases after that 
and their Reynolds number can drop below $Re_b$ before the next 
encounter takes place.} ($M_e\gtrsim M_f$). For even bigger 
planetesimals gas drag is quadratic and planetesimals
can be either in the shear- (small ones) or in the dispersion-dominated 
(large ones) regime. In the latter regime the dot-dashed 
line separates cases of equilibrium planetesimal random velocity
being higher or lower than $\Delta v_g$, see equations 
(\ref{eq:high_vel_cond}) and (\ref{eq:low_vel_cond}). When considering 
the population of small planetesimals 
we should always keep in mind that our discussion in \S 
\ref{subsect:high_vel}-\ref{subsect:Stokes} is valid only for 
planetesimals moving on almost Keplerian orbits. Thus, a condition 
$r_p\gtrsim r_{stop}$ must be satisfied which restricts the validity 
of our results to planetesimals bigger than $\sim 1$ m (see Figure 
\ref{fig:length_scales}). Smaller planetesimals largely follow 
the motion of the gas. Embryo's mass is not allowed to exceed 
$\eta^{-3/2}M_f$ because more massive protoplanetary cores excite 
supersonic random velocities of surrounding planetesimals.

As we have demonstrated in \S \ref{subsect:inclination}  
small shear-dominated planetesimals sediment into the  
geometrically thin subdisk (thinner than the embryo's Hill radius) 
near the midplane of the nebula. We can 
readily see from the Figure \ref{fig:earth} that in the terrestrial 
region this destiny awaits all planetesimals smaller than 
$\approx 100-200$ m in size (less than $\sim 10^{13}-10^{14}$ g 
in mass) in the vicinity of embryos more massive 
than about $10^{25}$ g. 
Near less massive embryos even larger planetesimals can belong to 
this dynamically cold population: only bodies bigger than about 
$1$ km can escape this fate near $10^{23}$ g mass embryo. Thus,
gas drag can have important effect even on $0.1-1$ km size 
planetesimals when the question of their interaction with 
the massive embryos is concerned.

\begin{figure}
\plotone{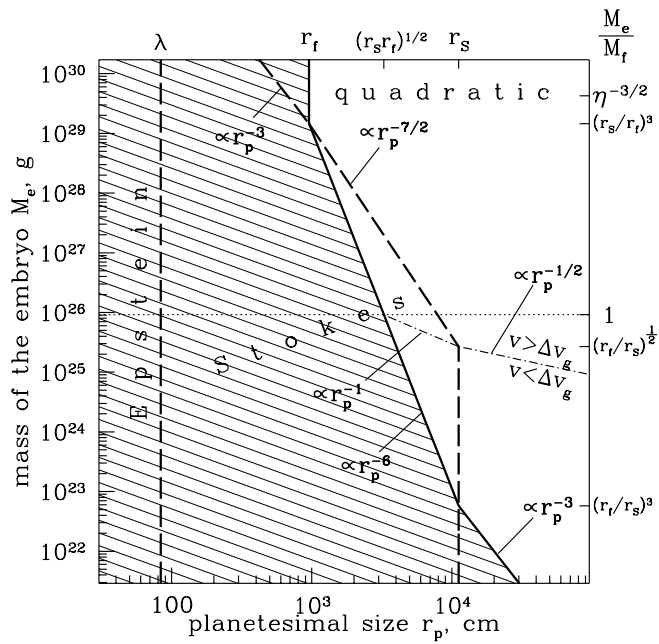}
\caption{
Same as Figure \ref{fig:earth} but for $a=5$ AU 
(giant planet region), where $\lambda\lesssim r_f\lesssim r_S$. 
\label{fig:jupiter}}
\end{figure}

Figure \ref{fig:jupiter} represents the separation of different 
gas drag and dynamical regimes at $5$ AU from the Sun
($\lambda\lesssim r_f\lesssim r_S$), 
corresponding to the giant planet region (roughly the semimajor
axis of Jupiter). Again, one should
keep in mind that planetesimal smaller than about $1$ m 
are tightly coupled to the gas at $5$ AU, see Figure 
\ref{fig:length_scales}. Evidently, the structure of the $M_e-r_p$
phase space is more complex in this part of the nebula. 
For example, planetesimals interacting with the embryo in the 
dispersion-dominated regime can now experience not only 
quadratic but also the Stokes drag. The reason for this is the 
lower Hill velocity $\Omega R_H$ for a given $M_e$ at $5$ AU which 
reduces the efficiency of planetesimal stirring by the embryo 
(see [\ref{eq:scat_rate}]) and diminishes the 
equilibrium planetesimal velocity. As a result, Reynolds number
can drop below $Re_b$ even for the dispersion-dominated planetesimals 
and drag can switch to the Stokes regime. 
Boundaries of different drag
regimes are computed using equations (\ref{eq:always_Stokes})
and (\ref{eq:Stokes_bnd}). Hatched region again represents 
$M_e$ and $r_p$ for which planetesimal scattering by the embryo
occurs in the shear-dominated regime (its boundary is determined
by equations [\ref{eq:dd_cond}], [\ref{eq:dd_cond1}], and 
[\ref{eq:Stokes_restr}]). One can see that this happens for  
planetesimals that are somewhat smaller than at 
$1$ AU: $r_p\lesssim 30$ m for $M_e\approx 10^{26}$ g
and $r_p\lesssim 100$ m for $M_e\approx 10^{23}$ g. This is 
because gas density rapidly drops with the distance from the Sun 
diminishing the strength of dissipation, and
this tendency cannot be counteracted by the longer synodic
period at $5$ AU (for a given $M_e$). Thus, somewhat
less planetesimal material will be concentrated in the vertically 
thin subdisk of small bodies at $5$ AU than at $1$ 
AU, but the difference is not very pronounced.

\begin{figure}
\plotone{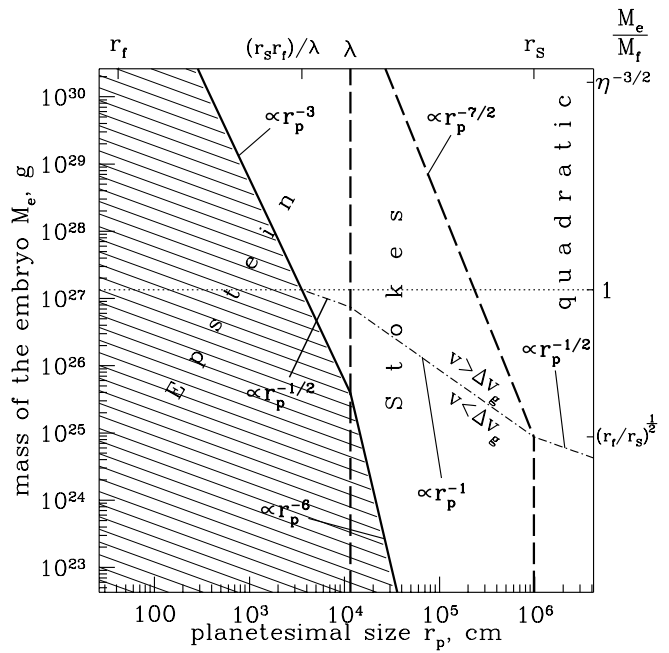}
\caption{
Same as Figure \ref{fig:earth} but for $a=30$ AU 
(region of ice giants), where $r_f\lesssim \lambda\lesssim r_S$. 
\label{fig:neptune}}
\end{figure}

Finally, Figure \ref{fig:neptune} displays the situation at $30$ AU
(roughly the semimajor axis of Neptune), in the region of
ice giants. Calculation of boundaries of different gas drag and 
dynamical regimes is performed using equations 
(\ref{eq:low_vel_cond}), 
(\ref{eq:Stokes_restr})-(\ref{eq:Stokes_bnd}), 
(\ref{eq:Epstein_restr}).
From Figure \ref{fig:length_scales} we can easily
see that the molecular mean free path $\lambda$ is very 
large in this
part of the nebula making the Epstein gas drag important for 
setting the boundary between the shear- and dispersion-dominated 
regimes. In this distant part of the nebula only bodies  
smaller than $\sim 10$ cm would be dynamically coupled to the gas.
Critical planetesimal size below which scattering is 
shear-dominated is $50-300$ m for embryos with masses 
$10^{23}-10^{27}$ g. This critical size is larger than at
$5$ AU because gas drag 
in the Epstein and Stokes regimes is 
more efficient than in the quadratic regime (for the same 
$r_p$ and $M_e$).


\subsection{Migration, gaps, and multiple embryos.}
\label{subsect:migration}


Conservation of Jacobi constant ensures that any change in
the epicyclic energy of planetesimal in the course 
of its scattering by the embryo is accompanied by the 
change in planetesimal semimajor axis. As 
a result, planetesimal surface density distribution gets
perturbed by the embryo
and a gap might form (Ida \& Makino 1993). 
For the shear-dominated planetesimals gap 
opening can be especially important, 
because in this dynamical regime 
planetesimal guiding centers are moved 
away from the embryo's orbit by $\sim R_H$ 
in a single passage at $h\approx R_H$, see 
(\ref{eq:close}) (Rafikov 2001). If this happens, 
accretion of planetesimals by the embryo can be severely 
affected (Rafikov 2003a).

At the same time, gas drag causes orbital decay 
of planetesimals between 
encounters --- they migrate towards the Sun.
This drift moves planetesimals which are located inside of
the embryo's orbit  further from it, facilitating the gap 
opening. On the other hand, on the outer side of the embryo's 
orbit gas drag causes planetesimals to drift {\it towards} the embryo,
and this tends to oppose the gap formation. As long as 
planetesimals are less massive than the embryo, they are all repelled
by scattering in the same way independent of their masses. 
Thus, the question of whether the gap outside of the embryo's 
orbit is cleared or not depends only on the 
damping timescale; there exists a critical size   
$r_{mig}$ such that planetesimals of this size initially at 
$h\sim R_H$ from the embryo, after being repelled by roughly 
$R_H$, can migrate back the same distance in a synodic period. 
Only these planetesimals would be 
accreted by an isolated embryo: planetesimals with 
$r_p\gtrsim r_{mig}$ are too big to be brought back to the 
embryo by the gas drag, and a gap forms preventing their 
further accretion. Planetesimals smaller than $r_{mig}$ 
migrate so fast that in a synodic period they cross the 
embryo's orbit and are lost to the inner disk. Embryo's 
accretion would then be rather inefficient as well.

These problems arise only for an {\it isolated} protoplanetary 
core. However, during the intermediate 
stages of planet formation
as a consequence of oligarchic growth (Kokubo \& Ida 1998)
there would be {\it many} embryos present in a disk at the 
same time. Their orbits should not be very widely separated:
even if it were the case initially, subsequent increase in 
the embryo masses caused by the accretion of planetesimals 
would make these separations not too large (see below) in terms 
of their Hill radii (because $R_H$ expands as embryo's mass 
increases, see [\ref{eq:Hill}]) and this is what is 
important for the dynamics. 

When such a ``crowded'' population of protoplanetary cores 
(which is a natural outcome of oligarchic growth) is present 
in the disk, gap formation is no longer an issue: although a 
particular embryo repels planetesimals and tends to open a gap, 
scattering by another nearby embryo pushes planetesimals back 
and spatially homogenizes them before they approach the 
first embryo again. Gap 
formation is thus suppressed and accretion can proceed almost 
uninhibited. On the other side of the problem, 
although small shear-dominated 
planetesimals migrate through the orbit of some
particular embryo because of the gas drag, there are many 
other embryos in the inner disk which scatter planetesimals 
back and forth, so that their inward drift looks more like a 
random walk through the nebula. In the course of such a 
diffusion through the protoplanetary disk planetesimals 
have a high chance of being accreted by one of the
many embryos (as long as planetesimals stay in the 
shear-dominated regime, see \S \ref{sect:accretion}). 

At a glance, it seems improbable that embryos can remain
on purely circular and uninclined orbits in a ``crowded'' 
configuration because at small radial separations
they would strongly scatter each other and very quickly excite 
large random velocities. However, one should remember that 
in the course of oligarchic growth most of the solid
mass is locked up in planetesimals and not in embryos.
This is true until $M_e$ reaches the 
isolation mass $M_{iso}$ defined as
\begin{eqnarray}
M_{iso}\approx M_\odot\left(4\pi \Sigma_p a^2/M_\odot\right)^{3/2}
\approx 6\times 10^{26}~g~a_{AU}^{3/4}
\label{eq:m_iso_cold}
\end{eqnarray}
if embryos grow predominantly by accretion of shear-dominated 
planetesimals and
\begin{eqnarray}
&& M_{iso}\approx M_\odot\left(4\pi \Sigma_p a^2/M_\odot\right)^{3/2}
\left(V/\Omega R_H\right)^{3/2}\nonumber\\
&& \approx 6\times 10^{26}~g~\left(V/\Omega R_H\right)^{3/2}a_{AU}^{3/4}
\label{eq:m_iso_hot}
\end{eqnarray}
if embryo growth is dominated by accretion of large 
dispersion-dominated planetesimals with typical random 
velocity $V$ (e.g. see Rafikov 2003b).  
Massive planetesimal population exerts {\it dynamical friction} 
on embryos transferring random energy of their 
epicyclic motion to planetesimals, which keeps embryo 
eccentricities small (see Kokubo \& Ida 1995). 
Planetesimal velocities, in turn, 
are damped by the gas drag\footnote{In the end, the energy 
of the embryo's epicyclic motion gets damped by the gas drag, with
planetesimal population acting as an intermediary.} which 
allows dynamical friction to continue being effective. 

Planetesimal disks have presumably contained both massive 
dispersion-dominated planetesimals and small shear-dominated 
bodies at the same time. We will assume that the latter comprise 
some fraction $\chi< 1$ of the total planetesimal surface density 
$\Sigma_p$. One can show (Stewart \& Ida 2000; Rafikov 2003c) that 
these two populations of planetesimals damp random velocity $v_e$ of 
protoplanetary cores at a rate
\begin{eqnarray}
&& \frac{dv_e^2}{dt}\approx -\Omega v_e^2
\left(\frac{M_{\odot}}{M_e}\right)^{1/3}
\frac{\Sigma_p a^2}{M_{\odot}}\nonumber\\
&& \times\left[\chi+(1-\chi)\left(
\frac{\Omega R_H}{V}\right)^{4}\right],
\label{eq:df_rate}
\end{eqnarray}
where second term represents the contribution of large 
dispersion-dominated planetesimals (we are dropping here all 
constant factors and the Coulomb logarithm), while the first term is due
to the small shear-dominated bodies\footnote{Effect of dynamical 
friction by the shear-dominated bodies can be obtained from that caused 
by the dispersion-dominated planetesimals 
by setting $v=\Omega R_H$ and reducing planetesimal surface density by 
a factor of $\chi$.}. Evidently, small bodies are more important
for ``cooling'' the cores than large planetesimals provided that
\begin{eqnarray}
\chi\gtrsim \left(\frac{\Omega R_H}{V}\right)^{4}.
\label{eq:dd_vs_sd}
\end{eqnarray}
Note that because $V\gtrsim \Omega R_H$ this inequality can
be fulfilled for rather small $\chi$ (e.g. for $\chi=0.02$ for
$V=3\Omega R_H$).
In the case opposite to (\ref{eq:dd_vs_sd}) dynamical friction 
on the protoplanetary cores is controlled by the large 
dispersion-dominated planetesimals. 

We demonstrate later in \S \ref{sect:accretion} that the relative 
importance of shear- or dispersion-dominated populations for 
the growth of mass of protoplanetary cores is determined by a 
condition different from (\ref{eq:dd_vs_sd}). Namely, protoplanetary 
cores grow predominantly through the accretion of small 
shear-dominated bodies whenever
\begin{eqnarray}
\chi\gtrsim p^{1/2}\left(\Omega R_H/V\right)^2,
\label{eq:acc_separ}
\end{eqnarray}
where parameter $p\ll 1$ is defined by equation (\ref{eq:p_def}). 
In this accretion regime one can also demonstrate that radial separation of 
embryos $h\approx R_H$ and their gravitational scattering 
has a discrete character: random velocities of embryos are strongly 
excited during their close approaches to each other, but dynamical 
friction tends to damp them before the next encounter occurs; this 
is similar to scattering of the shear-dominated planetesimals 
in the presence of gas drag (see 
\S \ref{subsect:high_vel}-\ref{subsect:inclination}). A particular 
embryo grows mainly by accretion of material (small planetesimals) 
from the annulus around its orbit (the so called ``feeding zone'') 
with the radial width $\approx R_H$ leading to isolation mass in the 
form (\ref{eq:m_iso_cold}). For a system of embryos to remain 
most of the time on uninclined\footnote{Analogous to the planetesimal case, the
shear-dominated scattering of embryos by embryos is very 
inefficient in exciting vertical velocities.}
and roughly circular orbits with radial 
spacing $h\approx R_H$, dynamical friction timescale $t_{df}$ 
must be shorter than the average time between the embryo encounters
$t_{syn}$ given by (\ref{eq:synodic}). Estimating $t_{df}$ from  
(\ref{eq:df_rate}) one can find that this is possible only for embryos 
with masses satisfying 
\begin{eqnarray}
&& M_e\lesssim  M_{e,cr}=M_\odot\left(\frac{\Sigma_p a^2}{M_\odot}\right)^{3/2}
\nonumber\\
&& \times\left[\chi+(1-\chi)\left(\frac{\Omega R_H}{V}\right)^4\right]^{3/2}.
\label{eq:constr1}
\end{eqnarray}
Note that if the whole planetesimal disk is shear-dominated, 
i.e. $\chi=1$, then embryos can stably stay on closely packed 
($h\approx R_H$) almost circular and uninclined orbits all the way until
they reach the isolation mass (modulo constant factors), compare with 
(\ref{eq:m_iso_cold}). However, if $\chi<1$ critical embryo mass 
goes down compared to $M_{iso}$.

\begin{figure}
\plotone{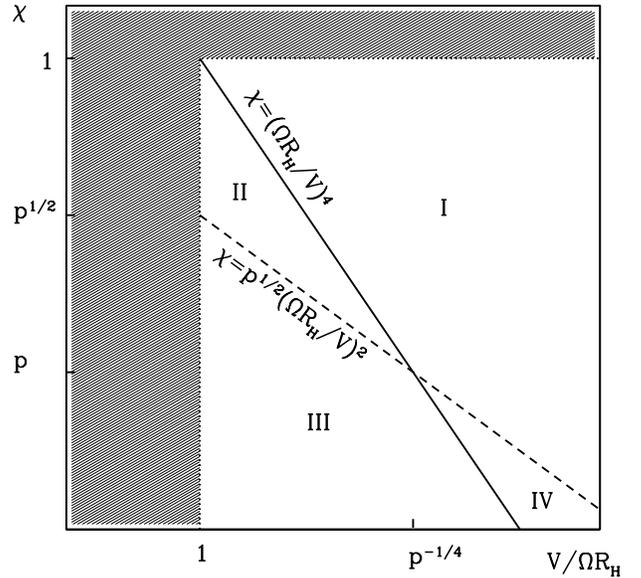}
\caption{
Different regimes of core accretion and dynamical friction in the 
presence of two populations of planetesimals --- shear- and 
dispersion-dominated. Separation of regimes is presented in terms 
of $V/\Omega R_H$ --- velocity of large planetesimals scaled by 
the embryo Hill velocity and $\chi$ --- planetesimal mass 
fraction locked up in small bodies. Solid and dashed lines 
correspond to conditions (\ref{eq:dd_vs_sd}) and 
(\ref{eq:acc_separ}).
\label{fig:acc_df}}
\end{figure}

Whenever the amount of solid mass in shear-dominated planetesimals 
is so small that (\ref{eq:acc_separ}) is violated and
$\chi\lesssim p^{1/2}(\Omega R_H/V)^2$,  core growth is determined mainly 
by the accretion of large dispersion-dominated planetesimals.
One can demonstrate in this regime that as a result of oligarchic growth 
radial separations of protoplanetary cores 
become $\approx V/\Omega$ (Ida \& Makino 1993; Kokubo \& Ida 1998) 
which is larger 
than $R_H$. Feeding zones of embryos are larger than in the 
shear-dominated case: dispersion-dominated planetesimals can
be accreted from within $V/\Omega$ in the radial direction (i.e. the 
width of the feeding zone is again roughly equal to the radial 
separation between the embryo orbits). Since orbits of embryos are 
well separated ($h\approx V/\Omega\gtrsim R_H$) their gravitational 
scattering is not so dramatic as in the case corresponding to 
(\ref{eq:acc_separ})
and eccentricity growth has a character of a random walk. Using 
equation (\ref{eq:distant_av}) and the fact that embryos encounter 
their neighbors roughly once every 
$t_{syn}(R_H/h)\approx t_{syn}(\Omega R_H/V)$ 
(where $t_{syn}$ is defined by [\ref{eq:synodic}]) we find growth 
rate of random velocity to be 
\begin{eqnarray} 
&& \frac{dv_e^2}{dt}\approx (\Omega a)^2
\left(\frac{M_e}{M_\odot}\right)^{2/3}
\left(\frac{R_H}{h}\right)^4 \times 
\Omega\left(\frac{M_e}{M_\odot}\right)^{1/3}\frac{h}{R_H}\nonumber\\
&& \approx \Omega(\Omega a)^2\frac{M_e}{M_\odot}
\left(\frac{\Omega R_H}{V}\right)^3. 
\label{eq:growth_rate}
\end{eqnarray}
For the system of embryos to be dynamically stable, the equilibrium 
value of random velocity $v_{e}$ (obtained by balancing scattering  
rate [\ref{eq:growth_rate}] with the damping due to the dynamical 
friction [\ref{eq:df_rate}]) has to be less than $V$ --- orbits of
cores should not cross. One can easily find that this is possible only 
provided that 
\begin{eqnarray}
&& M_e\lesssim M_{e,cr}=M_\odot\left(\frac{\Sigma_p a^2}{M_\odot}\right)^{3/2}
\left(\frac{V}{\Omega R_H}\right)^{3/2}\nonumber\\
&& \times\left[1-\chi+\chi\left(\frac{V}{\Omega R_H}\right)^4\right]^{3/2},
\label{eq:constr2} 
\end{eqnarray}
If planetesimal disk contains only dispersion-dominated planetesimals,
i.e. $\chi=0$, then population of embryos can be dynamical stable 
only until embryos reach corresponding $M_{iso}$ given by (\ref{eq:m_iso_hot}). 
In Figure \ref{fig:acc_df} we display different regimes 
of dynamical friction and planetesimal accretion as a function
of $\chi$ and $V/\Omega R_H$. 

Thus, whenever planetesimal disk is one-component (i.e. has either 
only shear-dominated or only dispersion-dominated planetesimals), it is the
isolation mass that sets a limit on the maximum core mass below which 
protoplanetary cores co-exist on closely spaced but dynamically 
cold orbits. However in the more realistic case of a disk consisting 
of both populations (i.e. $0<\chi<1$) dynamical stability is violated at 
masses smaller (sometimes much smaller) than $M_{iso}$, as equations 
(\ref{eq:constr1}) and (\ref{eq:constr2}) clearly demonstrate.  
 For instance, let us consider the region of 
giant planets at $a=5$ AU where $p\approx 10^{-3}$ according to (\ref{eq:p_def}). 
Assuming that most of the surface density is in large, 1-km size 
planetesimals, we find their 
random velocity $V\approx 2.2\Omega R_H$ to be set by quadratic gas 
drag (\ref{eq:bigger_offset}). Then, whenever the mass fraction of small
shear-dominated planetesimals forming almost two-dimensional subdisk 
is $\chi\lesssim 6\times 10^{-3}$, both core accretion rate and dynamical 
friction are set only by large planetesimals. For such small $\chi$ the 
system of embryos is dynamically stable until $M_e$ reaches the 
isolation mass (\ref{eq:m_iso_hot}), see (\ref{eq:constr2}). For 
$6\times 10^{-3}\lesssim\chi\lesssim 4\times 10^{-2}$ situation is 
different: dynamical friction acting on cores is still determined by large 
planetesimals but core accretion rate is now set by the abundance of 
small bodies. Orbits of embryos 
accreting small bodies tend to get closely packed\footnote{When $\chi$ 
passes through the value $p^{1/2}(\Omega R_H/V)^2$ equilibrium radial 
separation between the embryo orbits discontinuously changes from $V/\Omega$
to $R_H$.} and this increases gravitational scattering between them. 
As a result, dynamical instability sets in at a considerably 
smaller mass $M_{e,cr}\approx M_{iso}(\Omega R_H/V)^6$, where $M_{iso}$ 
is now given by (\ref{eq:m_iso_cold}); for $V=2.2\Omega R_H$ one finds
$M_{e,cr}\approx 0.01 M_{iso}$. Finally, when $\chi\gtrsim 
4\times 10^{-2}$ both dynamical friction and core accretion rates are 
determined by small shear-dominated bodies and $M_{e,cr}\approx M_{iso}\chi^{3/2}$,
($M_{iso}$ is again given by [\ref{eq:m_iso_cold}]); only when $\chi=1$
can instability be postponed until $M_e$ reaches $M_{iso}$. 

From this example one can see that the situation in two-component 
planetesimal disks is very different from that in one-component disks. 
Even small admixture  of shear-dominated planetesimals 
can completely change the dynamics of a system of protoplanetary 
cores. Thus, it becomes especially important to know the exact 
value of $\chi$ and to follow its evolution in time.
Table 1 briefly summarizes our findings by delineating the conditions 
under which shear-dominated or dispersion-dominated 
planetesimals control core accretion and dynamical friction.

As cores grow, their Hill radii increase and feeding zones overlap 
leading to occasional mergers of cores. This keeps their orbital 
separations from becoming too small but does not lead to dynamical 
instability because of the dynamical friction. Thus, the overall 
picture of oligarchic growth described above is not affected by 
mergers of embryos. However, as soon as $M_e$ increases 
beyond the threshold given by (\ref{eq:constr1}) or (\ref{eq:constr2}), 
dynamical friction can no longer keep embryos on kinematically cold 
orbits, their eccentricities and inclinations start to grow, and 
cores finally switch into the dispersion-dominated regime with respect 
to each other. Similar effect has been observed by Kokubo \& 
Ida (1995) in N-body simulations, although they were dealing 
with the gas-free environment which made dynamical friction 
less effective since planetesimals were dynamically hot. In this 
work we are not going to follow this more violent stage of 
planetesimal disk evolution.


\section{Lower limit on planetesimal velocity.}
\label{sect:lower}


We have demonstrated in \S \ref{subsect:high_vel} that 
planetesimal random velocity in the shear-dominated case 
rapidly decays as a result of gas drag after scattering by 
the embryo. If the damping time $t_d$ is much shorter than the
time between encounters, planetesimal velocity 
right before the next approach to the embryo would  
essentially be zero [$\propto \exp(-t_{syn}/t_d)$ 
to be more exact]. 
As a result, any subdominant sources of stirring which 
would normally be negligible 
become important in maintaining random motions at 
a finite value. These effects then 
determine the floor below which planetesimal random velocity 
cannot drop. Here we identify two such effects ---
scattering by the distant embryos and stirring by the large 
dispersion-dominated planetesimals --- which determine the 
minimum horizontal and vertical random velocities of 
planetesimals respectively. We consider these two
processes separately.

As we demonstrated in \S \ref{subsect:migration},  
proto-Solar nebula should contain a population of 
dynamically cold embryos separated by roughly $R_H/N$ --- we 
parametrize the uncertainty in the radial separation of cores 
by the number of embryos $N$ per $R_H$ in radius; if cores grow 
mainly through accretion of shear-dominated planetesimals 
$N\approx 1$, while if they mostly accrete dispersion-dominated 
bodies $N\approx (\Omega R_H/V)<1$, see \S \ref{subsect:migration}. 
Surface number density of embryos is $\approx N/(2\pi a R_H)$.

Distant embryos scatter a given planetesimal quite frequently
because both the velocity of incoming embryos (which is determined
by the shear in the disk) and their number increase 
linearly with the radial separation $h$. As a result, although 
scattering by the nearest embryos is discrete,  
scattering by the cores more distant than some critical
$h_c$ should be considered as a continuous process, similar to
the scattering in the dispersion-dominated regime. To determine 
$h_c$ we notice that scattering switches from 
the discrete to continuous mode when the average time between 
the passages of embryos separated from a given planetesimal by less
than $h_c$ becomes shorter than the typical timescale 
on which planetesimal velocity would evolve otherwise. For 
planetesimals with $r_p\gtrsim r_{stop}$ this typical timescale 
is the gas damping timescale\footnote{Note that for planetesimals 
smaller than $r_{stop}$ gas damping time is shorter than 
$\Omega^{-1}$. However the duration of the gravitational interaction 
with the distant embryo is always $\sim\Omega^{-1}$ meaning that
the typical velocity evolution timescale for planetesimals with
$r_p\lesssim r_{stop}$ is not $t_d$ but rather $\Omega^{-1}$.
This, however, is only important for very small planetesimals, 
$r_p\lesssim 1$ m (see Figure \ref{fig:length_scales}), 
which are not covered by this study anyway.} $t_d$. 

The rate $\Gamma(h_c)$ at which embryos with $|h|<h_c$ 
pass a particular planetesimal due to the shear in the disk is
\begin{eqnarray}
\Gamma(h_c)=\frac{3\Omega}{2}\frac{N}{2\pi a R_H}
\int\limits_{-h_c}^{h_c}|h|dh=\Omega\frac{3}{4\pi}
\frac{N}{a R_H}h_c^2.
\label{eq:rate}
\end{eqnarray} 
Boundary between the discrete and continuous scattering is 
given by $\Gamma(h_c)t_d\approx 1$, meaning that
\begin{eqnarray}
h_c\approx \left[\frac{4\pi}{3}\frac{a R_H}{N}
(\Omega t_d)^{-1}\right]^{1/2}=R_H\left(\frac{4\pi}{3N}
\frac{t_{syn}}{t_d}\right)^{1/2}.
\label{eq:h_c}
\end{eqnarray}
Thus, $h_c\gg R_H$ for $N\lesssim 1$ and $t_d\ll t_{syn}$. 

Scattering by the embryos with $|h|\gg h_c$ occurs so
frequently compared to the gas damping timescale that 
the random component of scattering averages to zero 
and eccentricity stirring is given by  (\ref{eq:distant_av}). 
We find that
\begin{eqnarray}
&& \frac{de^2}{dt}\approx\frac{3\Omega}{2}\frac{N}{2\pi a R_H}\times 2
\int\limits_0^{h_c}\langle\Delta (e^2)\rangle |h|dh
\nonumber\\
&& =5\frac{3N}{4\pi}\Omega N\left(\frac{R_H}{a}\right)^3
\left(\frac{R_H}{h_c}\right)^2.
\label{eq:stir_dist}
\end{eqnarray}
Balancing this stirring by the gas damping $de^2/dt=-e^2/t_d$
and using (\ref{eq:h_c}) we find the equilibrium value 
of eccentricity
\begin{eqnarray}
e_{min}\approx \sqrt{5}\frac{3N}{4\pi}\frac{R_H}{a}
\frac{t_d}{t_{syn}}\sim \frac{R_H}{a}\left(\frac{R_H}{h_c}\right)^2.
\label{eq:eq_ecc}
\end{eqnarray}
Using (\ref{eq:damp_quad})-(\ref{eq:damp_Epstein}) 
we evaluate 
\begin{eqnarray}
&& e_{min}\approx 0.04~ N\frac{R_H}{a}
\left(\frac{M_e}{10^{25}\mbox{g}}\right)^{1/3}
\frac{r_p}{10~\mbox{m}}a_{AU}^{5/4},
\label{eq:eq_ecc_quad}\\
&& e_{min}\approx 0.03~ N\frac{R_H}{a}
\left(\frac{M_e}{10^{25}\mbox{g}}\right)^{1/3}
\left(\frac{r_p}{10~\mbox{m}}\right)^{2}
\left(\frac{a_{AU}}{5}\right)^{-5/4},
\label{eq:eq_ecc_Stokes}\\
&& e_{min}\approx 0.05~ N\frac{R_H}{a}
\left(\frac{M_e}{10^{25}\mbox{g}}\right)^{1/3}
\frac{r_p}{10~\mbox{m}}
\left(\frac{a_{AU}}{30}\right)^{3/2},
\label{eq:eq_ecc_Epstein}
\end{eqnarray}
for quadratic, Stokes, and Epstein drag regimes respectively.
These estimates imply that the minimum horizontal random 
velocities of small planetesimals are below
the Hill velocity, as they should be.
For 10 m planetesimals stirred by  $10^{25}$ g embryos
$e_{min}$ corresponds to velocities of the order
of 1 m s$^{-1}$ in the inner part of the 
proto-Solar nebula, dropping to $\approx 0.5$ m s$^{-1}$ 
at 30 AU.

At the same time, scattering by distant embryos 
cannot maintain the inclinations of 
small planetesimals at a finite level. Excitation of the vertical 
velocity by an encounter with an embryo separated even 
by $R_H$ from planetesimal is weakened 
compared to the excitation of horizontal 
velocity by the geometric factor $ia/R_H\ll 1$. As a result, 
the growth rate of inclination due to the 
embryo scattering scales 
as $i^2$ (exactly like gas drag) and for $t_d\lesssim t_{syn}$
gas drag unconditionally dominates. We now consider if 
stirring by planetesimals (and not embryos)
can keep nonzero inclinations of small bodies.

Gas drag acting on planetesimals bigger than 
0.1--1 km (depending on the location in the nebula, 
see \S \ref{subsect:separation}) is too weak to 
prevent them from staying in the dispersion-dominated regime with 
respect to embryos. Gravitational interaction of these massive 
planetesimals with small bodies is certain to take place in the 
dynamically ``hot'' regime (because for the same physical 
velocity the Hill radius for the planetesimal-planetesimal 
scattering is much smaller than $R_H$ for the embryo-planetesimal 
scattering).

Rafikov (2003c) has demonstrated that velocity excitation
by planetesimals sensitively depends on the 
planetesimal mass spectrum. For a given differential 
surface number 
density distribution of planetesimal masses $d{\cal N}/dm$ the 
inclination stirring can be written as (Rafikov 2003c)
\begin{eqnarray} 
\frac{di^2}{dt}\approx\Omega  
\left(\frac{\Omega a}{V}\right)^2\frac{a^2}{M_\odot^2}
\int \frac{d{\cal N}(m)}{dm}m^2dm,
\label{eq:stir_big}
\end{eqnarray}
where $V$ is the random velocity of large dispersion-dominated 
planetesimals (which we for simplicity set constant,  
independent of $m$).
Shear-dominated planetesimals are not efficient at
velocity excitation (but they matter for dynamical friction). 
Because of the stirring by embryos $V$ 
should be some multiple of $\Omega R_H$, 
but owing to the action of gas drag (and 
planetesimal dynamical friction) it is 
not higher than several $\Omega R_H$, see equations 
(\ref{eq:smaller_offset}), (\ref{eq:vel_Stokes}), and 
(\ref{eq:vel_Epstein}).

\begin{figure}
\plotone{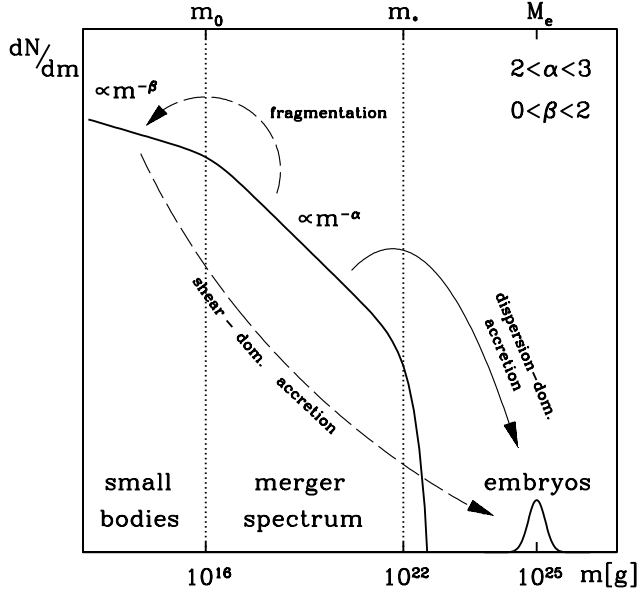}
\caption{
Schematic representation of the mass spectrum of solid bodies 
in the disk (thick solid curve). Three distinct parts of the 
spectrum can be singled out: small bodies ($d{\cal N}/dm\propto 
m^{-\beta}, ~0<\beta<2$, most of the mass is in biggest objects),
merger spectrum  ($d{\cal N}/dm\propto 
m^{-\alpha}, ~2<\alpha<3$, most of the mass is in smallest objects),
and a set of embryos detached from continuous mass distribution
(see \S \ref{sect:lower}).
Arrows describe different mass transfer routes between the components
of the mass spectrum: solid arrow is for the direct accretion of large 
(dispersion-dominated) planetesimals by embryos, while dashed 
arrows describe fragmentation of planetesimals with further 
shear-dominated accretion by embryos. See \S \ref{sect:discus} 
for more details.
\label{fig:mass_spectrum}}
\end{figure}

Numerical simulations very often produce
planetesimal mass spectra such that 
$d{\cal N}/dm\propto m^{-\alpha}$ with 
$\alpha\approx 2.5$ within a wide range of masses
(Kokubo \& Ida 1996). In a disk with such spectrum 
most of the mass is concentrated at the lower end of 
distribution while most of the stirring 
is done by its upper end (Rafikov 2003c). 
We will assume that planetesimal mass spectrum 
has this form for $m_0\lesssim m_p\lesssim 
m_\star$, where we somewhat arbitrarily take $m_0=10^{16}$ g
and $m_\star\approx 10^{22}$ g (roughly 1-km and 100-km size 
planetesimals); see Figure \ref{fig:mass_spectrum} for a 
schematic picture of the assumed size distribution. 
Excitation of inclination by the largest planetesimals 
($m_p\approx m_\star$, which are still much smaller than 
the embryos) can be expressed in 
terms of $m_0$, $m_\star$, and planetesimal surface mass 
density $\Sigma_p$ (dominated by planetesimals with 
$m_p\approx m_0$) roughly as (Rafikov 2003c)
\begin{eqnarray} 
\frac{di^2}{dt}\approx\Omega  
\left(\frac{\Omega a}{V}\right)^2\frac{m_0 \Sigma_p a^2}
{M_\odot^2}\left(\frac{m_\star}{m_0}\right)^{3-\alpha}.
\label{eq:stir_big1}
\end{eqnarray}
In writing down this expression we 
have assumed that large planetesimals are
numerous enough to give rise to a continuous rather
than discrete stirring. Balancing (\ref{eq:stir_big1}) 
with $-i^2/t_d$ we find the equilibrium value 
of inclination:
\begin{eqnarray}
i_{min}\approx\frac{\Omega a}{V}\left[\Omega t_d\frac{m_0\Sigma_pa^2}
{M_\odot^2}\left(\frac{m_\star}{m_0}\right)^{3-\alpha}
\right]^{1/2}.
\label{eq:eq_incl}
\end{eqnarray} 
Evaluating this expression for $V=3\Omega R_H$, $M_e=10^{25}$ g,
$\Sigma_p$ given by (\ref{eq:MMSN}), $t_d$ given by 
(\ref{eq:damp_quad})-(\ref{eq:damp_Epstein}), and our adopted
values of $m_0$, $m_\star$, and $\alpha=2.5$ we find that
\begin{eqnarray}
&& i_{min}\approx 10^{-4}\frac{R_H}{a}
\left(\frac{r_p}{10~\mbox{m}}\right)^{1/2}
a_{AU}^{7/8},
\label{eq:i_eq_quad}\\
&& i_{min}\approx 10^{-4}\frac{R_H}{a}
\frac{r_p}{10~\mbox{m}}
\left(\frac{a_{AU}}{5}\right)^{-3/8},
\label{eq:i_eq_Stokes}\\
&& i_{min}\approx 3\times 10^{-4}\frac{R_H}{a}
\left(\frac{r_p}{10~\mbox{m}}\right)^{1/2}
\frac{a_{AU}}{30},
\label{eq:i_eq_Epstein}
\end{eqnarray}
for corresponding gas drag regimes.
These values of inclination correspond to very small 
vertical velocities of planetesimals, $\sim 
1$ cm s$^{-1}$ in the vicinity of $10^{25}$ g protoplanetary
cores at $30$ AU. Evidently, since $i_{min}\ll e_{min}$ (see 
[\ref{eq:eq_ecc_quad}]-[\ref{eq:eq_ecc_Epstein}]) this 
type of stirring would not be 
able to affect minimum eccentricities of planetesimals. 

Scattering by large planetesimals would vertically 
perturb not only small bodies but also embryos. 
Balancing the stirring rate (\ref{eq:stir_big1}) with 
the dynamical friction rate (\ref{eq:df_rate}), we 
find that the minimum inclination of the embryo $i_{e,min}$ is about
\begin{eqnarray}
&& i_{e,min}\approx \frac{R_H}{a}\left(
\frac{V}{\Omega R_H}\right)
\left[\frac{m_0}{M_e}\left(\frac{m_\star}
{m_0}\right)^{3-\alpha}\right]^{1/2}\nonumber\\
&& \times\left[1+\chi\left(
\frac{V}{\Omega R_H}\right)^4\right]^{-1/2}\\
&& \approx 
10^{-3}~\frac{R_H}{a}\left(
\frac{V}{\Omega R_H}\right)
\left(\frac{M_e}{10^{25}~\mbox{g}}\right)^{-1/2}\left[1+\chi\left(
\frac{V}{\Omega R_H}\right)^4\right]^{-1/2},\nonumber
\label{eq:embryo_min_incl}
\end{eqnarray}
provided that the time between encounters with 
large planetesimals (which dominate vertical stirring) 
is shorter than $t_{df}$ 
(for continuous approximation to hold). 
As a bottom 
line, we may conclude that 
 scattering by large planetesimals  keeps
inclinations of small bodies relative to 
protoplanetary cores at the level of 
$\sim 10^{-3}(R_H/a)$, most likely via the 
stirring of cores. 

There are other possibilities for
maintaining inclinations of shear-dominated planetesimals 
at some minimum level. 
One of them is the gravitational 
scattering between small bodies themselves which  
transfers the energy of random motion from horizontal
into the vertical direction. 
This is likely not to be important because 
dynamical relaxation of small bodies  is very slow. 
Another possibility is a gravitational instability in the 
thin layer (P. Goldreich, private communication) 
which may excite random velocities of small 
constituent bodies. Assuming that a fraction $\chi$ of
solid mass is locked up in small shear-dominated 
bodies, one can find that instability 
would operate if the planetesimal velocity dispersion is below
$\chi\pi G\Sigma_p/\Omega\approx 30$ cm s$^{-1}~\chi$ 
(usual Toomre criterion), and this is still below 
$e_{min}\Omega a$ even for $\chi=1$.
Finally, inelastic collisions between small planetesimals
can heat the disk vertically, and we elaborate more 
on this in the next section.


\section{Inelastic collisions.}
\label{subsect:inelastic}


Inelastic collisions between planetesimals act as efficient 
source of damping. The escape speed 
from the surface of a 100 m body is about 
10 cm s$^{-1}$ and
planetesimals are typically moving with higher 
velocities   
(see [\ref{eq:eq_ecc_quad}]-[\ref{eq:eq_ecc_Epstein}]),
which means that (a) they would lose a lot of energy
in high-energy collisions, and (b) gravitational 
focusing is unimportant and collision cross-section is
almost equal to the geometric cross-section of colliding bodies. 
Assuming 
that planetesimals lose a fraction $\sim 1$ of their energy 
when they collide, we estimate the velocity damping in inelastic
collisions to be given by 
\begin{eqnarray}
\frac{dv^2}{dt}\approx -\Omega v^2\frac{\chi\Sigma_p}{\rho_p r_p}
\frac{v_h}{v_z},
\label{eq:inel}
\end{eqnarray}
where we have again assumed that only a fraction $\chi$ of 
solid mass is in the shear-dominated planetesimals. 
From this expression it is clear that inelastic collisions
lead to exponential damping of velocity 
(if $v_h/v_z\approx const$) on a timescale 
\begin{eqnarray}
&& t_{inel}\equiv \Omega^{-1}\chi^{-1}
(\rho_p r_p/\Sigma_p)(v_z/v_h)\nonumber\\
&& \approx 5~\mbox{yr}~\chi^{-1}(v_z/v_h)\frac{r_p}{10~\mbox{m}}a_{AU}^3.
\label{eq:t_inel} 
\end{eqnarray}

Importance of inelastic collisions is judged by comparing
$t_{inel}$ with $t_d$ --- damping time due to the gas drag.
Using equations (\ref{eq:damp_quad})-(\ref{eq:damp_Epstein})
we find 
\begin{eqnarray}
&& t_d/t_{inel}\approx 0.25~\chi\left(v_h/v_z\right)a_{AU}^{-1/4},
\label{eq:timesc_rat_quad}\\
&& t_d/t_{inel}\approx 0.1~\chi\left(v_h/v_z\right)
\left(\frac{r_p/\lambda}{10}\right),
\label{eq:timesc_rat_Stokes}\\
&& t_d/t_{inel}\approx 0.01~\chi\left(v_h/v_z\right),
\label{eq:timesc_rat_Epstein}
\end{eqnarray}
for quadratic, Stokes, and Epstein gas drag regimes.
These estimates clearly demonstrate that $t_d\lesssim t_{inel}$
whenever $v_z\sim v_h$ ($i\sim e$), meaning that inelastic collisions are 
unimportant when planetesimal velocities are roughly isotropic. 
This is always true in the dispersion-dominated regime 
allowing us to neglect inelastic collisions in this 
case.

Situation is different in the shear-dominated regime where one 
can easily have $v_z\ll v_h$ (see \S \ref{sect:lower}). 
Let's assume that planetesimal inclination right 
before the encounter with some embryo is very low. Right 
after the scattering event inclination remains roughly the same 
(see [\ref{eq:close_incl}]) while 
eccentricity goes up to $\sim (R_H/a)\gg i_{min}$; 
as a consequence, $t_{inel}$ immediately becomes much 
shorter than $t_d$ and inelastic collisions suddenly 
become more important than gas drag. 
However, this is a very transient stage because inelastic collisions
not only dissipate energy but also isotropize
the velocities of colliding bodies. If planetesimals 
were colliding like rigid balls, bouncing off after collision, 
one would expect roughly isotropic recoil velocities.
Consequently,  one would expect $v_z\approx v_h$  
after every planetesimal has experienced a single physical collision, 
i.e. after time $t_{inel}$ has passed since the scattering
by the embryo. This immediately reduces the importance 
of inelastic collisions and
makes gas drag more important again for the random 
velocity dissipation shortly after scattering has occurred, see 
(\ref{eq:timesc_rat_quad})-(\ref{eq:timesc_rat_Epstein}). 

After that, according to (\ref{eq:incl_decay}), 
inclination decays  slower
than eccentricity does meaning that the role 
of inelastic collisions 
keeps decreasing. If the time interval 
between the successive encounters with
the embryo is long enough, the eccentricity decay will stop at 
the asymptotic value $e_{min}$ (of course, with occasional 
oscillations due to scattering by distant embryos) while 
the inclination would continue to decay further until it reaches 
$i_{min}\ll e_{min}$. At this stage $v_z/v_h$ goes down and 
physical collisions can again start occurring quite 
frequently; however, the imminent isotropization of velocities 
after every collision limits their importance (in comparison with 
that of the gas drag) only to short periods of time. As a result, 
the timescale of velocity damping between encounters with 
embryos should still be very close to the gas damping time 
scale $t_d$.

Physical collisions might affect the determination of $i_{min}$ 
since they can be effective at pumping the energy of horizontal
motions into vertical ones. The minimum velocity anisotropy 
$v_z/v_h$ they provide can be determined from the condition 
$t_d/t_{inel}\approx 1$ and turns out to be roughly 
$0.25\chi$, $0.1\chi$, and $0.01\chi$ for typical
parameters of quadratic, Stokes, and Epstein drag regimes,
see equations (\ref{eq:timesc_rat_quad})-(\ref{eq:timesc_rat_Epstein}). 
Scattering by distant embryos maintains horizontal random
velocities of small planetesimals at the level given by 
(\ref{eq:eq_ecc_quad})-(\ref{eq:eq_ecc_Epstein}), and this degree 
of anisotropy results in $i_{min}\approx (v_z/v_h)e_{min}$
which for $10$ m planetesimals  is about $0.01\chi(R_H/a)$ at 
1 AU and $5\times 10^{-4}\chi(R_H/a)$ at $30$ AU. 
Comparing this with (\ref{eq:i_eq_quad})-(\ref{eq:i_eq_Epstein}) and 
(\ref{eq:embryo_min_incl}) we may 
conclude that for not too small values of $\chi$ ($\gtrsim 0.1$) 
inelastic collisions are important for setting the minimum 
value of relative embryo-planetesimal inclination in the presence of gas 
drag in quadratic or Stokes regimes (in the inner parts of the proto-Solar
nebula, inside $5-10$ AU). Whenever 
shear-dominated planetesimals are affected by the gas 
drag in the Epstein regime, inelastic collisions can compete with 
stirring by large planetesimals only for $\chi \approx 1$. 

This discussion has assumed rather idealized model of 
planetesimal collisions (rigid balls). In reality 
high-energy impacts are likely to be 
catastrophic, leading to the disruption of participant 
bodies. Then the amount of kinetic energy transferred 
into vertical motions and the degree of isotropization
would be determined by the ejection velocities of resulting 
debris. Observations of velocity dispersions in collisional 
families of asteroids (Zappala \etal 1996) and 
results of numerical simulations (Michel \etal 2003) 
indicate that ejection velocities are considerably 
smaller than the initial planetesimal velocities. This 
slows down the velocity isotropization compared to 
the case of collisions of ``hard balls''. 
Nevertheless, we do not expect this to seriously 
change the general picture outlined before.  
Besides, it is not at all clear how anisotropic would 
ejection velocities be in the case of 
planetesimals (10-100 m in size, 
strength-dominated fragmentation) colliding at 
several tens of m s$^{-1}$,
since collisions between 10-km asteroids 
(gravity-dominated fragmentation),
on which information exists, occur at relative 
speeds of several km s$^{-1}$.


\section{Accretion of low-energy planetesimals.}
\label{sect:accretion}


Protoplanetary embryos grow by accreting planetesimals. Accretion
in differentially rotating disks is intrinsically complicated because
of the three-body gravitational interaction of the two 
merging bodies and the central mass. One can however greatly simplify 
this problem by treating accretion as a two-body process while
approximately taking into account three-body effects by limiting from below
planetesimal approach velocity relative to the embryo by 
$\Omega R_H$. This is a direct consequence of the shear
present in the disk. 

With this in mind we can write a rather general formula for the 
protoplanetary mass accretion rate:
\begin{eqnarray}
&& \frac{dM_e}{dt}\approx n_p m_p R_e^2 v_a\left(1+\frac{v_{esc}^2}{v_a^2}\right)
\nonumber\\
&& \approx\Omega\Sigma_p R_H^2\frac{R_e}{R_H}\frac{v_a}{v_z}
\left(\frac{\Omega R_H}{v_a}\right)^2,
\label{eq:accr_rate_3D}
\end{eqnarray}
where $R_e$ is the physical size of the embryo and 
the last equality holds for $v_a\ll v_{esc}$. 
Dimensionless physical size of the core $p$ 
relative to its Hill radius is independent of the core 
mass but varies with the distance from the Sun:
\begin{eqnarray}
p\equiv \frac{R_e}{R_H}= \left(\frac{3}{4\pi}\frac{M_\odot}{\rho_p a^3}
\right)^{1/3}\approx 5.2\times 10^{-3}a_{AU}^{-1}.
\label{eq:p_def}
\end{eqnarray}
Clearly, $R_e\ll R_H$ and $p\ll 1$.

Scattering of planetesimals by the embryo in the 
dispersion-dominated regime always tends to maintain $v_z\approx v
\approx v_a\gtrsim \Omega R_H$, thus
\begin{eqnarray}
\frac{dM_e}{dt}\approx \Omega p\Sigma_p R_H^2
\left(\frac{\Omega R_H}{v}\right)^2.
\label{eq:accr_rate_3D_dd}
\end{eqnarray}

In the shear-dominated regime approach velocity $v_a$ is almost
independent of planetesimal eccentricity or inclination and is 
$\approx\Omega R_H$. At the same time, vertical velocity of 
planetesimals $v_z$ setting the thickness of planetesimal disk 
is smaller than $\Omega R_H$, and we find that 
\begin{eqnarray}
\frac{dM_e}{dt}\approx \Omega p\Sigma_p R_H^2
\frac{\Omega R_H}{v_z}.
\label{eq:accr_rate_3D_sd}
\end{eqnarray}

Derivation of equation (\ref{eq:accr_rate_3D}) has 
implicitly assumed that the maximum
impact parameter at infinity with which planetesimal 
can be accreted by the 
embryo $R_e\sqrt{1+v_{esc}^2/v_a^2}$ is smaller than the 
planetesimal disk thickness $v_z/\Omega$. In the shear-dominated 
case, when $v_a\sim\Omega R_H$, this 
sets a limit on the vertical velocity of planetesimals at which
their accretion can still be described by equation 
(\ref{eq:accr_rate_3D}):
\begin{eqnarray}
v_z\gtrsim v_{z,cr}\equiv p^{1/2}\Omega R_H 
\approx 
0.07~\Omega R_H~a_{AU}^{-1/2}
\label{eq:limit}
\end{eqnarray} 
(Greenberg \etal 1991; Dones \& Tremaine 1993).  
Equation (\ref{eq:accr_rate_3D_sd}) 
is applicable only for $v_z\gtrsim v_{z,cr}$.
Whenever $v_z\lesssim v_{z,cr}$ planetesimal disk is 
very thin and embryo can accrete the whole vertical column
of material it encounters (eccentricity is only restricted 
to be smaller than $R_H/a$). In this case one can easily show that
\begin{eqnarray}
&& \frac{dM_e}{dt}\approx \Omega\Sigma_p R_e R_H\left(1+\frac{v_{esc}^2}{v_a^2}
\right)^{1/2}\nonumber\\
&& \approx \Omega p^{1/2}\Sigma_p R_H^2.
\label{eq:accr_rate_2D}
\end{eqnarray}
This is the highest possible accretion rate of planetesimals 
by the protoplanetary core that can be achieved in 
the planetesimal disk.

It is possible that most of the planetesimal mass has been concentrated 
in bodies with sizes of about $1-10$ km  (merger products in Figure
\ref{fig:mass_spectrum}) which interact with embryos 
in the dispersion-dominated regime. 
According to (\ref{eq:accr_rate_3D_dd}) the growth timescale of
the embryo due to the accretion of these bodies is 
\begin{eqnarray}
&& t_{g,dd}\approx \Omega^{-1}\left(\frac{M_e}{M_\odot}\right)^{1/3}
\frac{M_\odot}{\Sigma_p a^2}\frac{R_H}{R_e}
\left(\frac{v}{\Omega R_H}\right)^2\nonumber\\
&& \approx 10^5~\Omega^{-1}
\left(\frac{M_e}{10^{25}\mbox{g}}\right)^{1/3}
\left(\frac{v}{\Omega R_H}\right)^2 a_{AU}^{1/2}.
\label{eq:t_grow_dd}
\end{eqnarray}
Using equations (\ref{eq:smaller_offset}), (\ref{eq:vel_Stokes}), and 
(\ref{eq:vel_Epstein}) we find that the time needed to
build $10^{25}$ g protoplanetary core by accretion of such 
dispersion-dominated planetesimals is $\approx 10^5$ yr at 1 AU,
$\approx 4\times 10^6$ yr at 5 AU, and 
$\approx 3\times 10^7$ yr at 30 AU.

At the same time, as we hypothesized in \S \ref{subsect:migration} 
a fraction $\chi<1$ of planetesimal mass 
could have been in small bodies which are shear-dominated with 
respect to embryos. This mass is concentrated in a vertically thin 
population and comparing (\ref{eq:i_eq_quad})-(\ref{eq:i_eq_Epstein}) 
\& (\ref{eq:embryo_min_incl})
with (\ref{eq:limit}) we find vertical
velocity of planetesimals in the subdisk to be smaller than 
$v_{z,cr}$. Thus, small shear-dominated bodies can be
very efficiently consumed  by embryos, and we find using 
(\ref{eq:accr_rate_2D}) that the embryo's growth timescale 
due to their accretion is
\begin{eqnarray}
&& t_{g,sd}\approx \chi^{-1}\Omega^{-1}\left(\frac{M_e}{M_\odot}\right)^{1/3}
\frac{M_\odot}{\Sigma_p a^2}\left(\frac{R_H}{R_e}\right)^{1/2}
\nonumber\\
&& \approx 7\times 10^3 \chi^{-1}\Omega^{-1}
\left(\frac{M_e}{10^{25}\mbox{g}}\right)^{1/3}.
\label{eq:t_grow_sd}
\end{eqnarray}
This is $\approx 10^3\chi^{-1}$ yr at 1 AU, $\approx 10^4\chi^{-1}$ 
yr at 5 AU, and 
$\approx 2\times 10^5\chi^{-1}$ yr at 30 AU. Thus, even if 
only about $1\%$ of the mass in solids is 
locked up in the 
shear-dominated planetesimals ($\chi=10^{-2}$) they would 
still dominate the accretion rate of the embryo
because $t_{g,sd}\lesssim t_{g,dd}$! Comparing 
(\ref{eq:accr_rate_3D_dd}) and (\ref{eq:accr_rate_2D}) with
$\Sigma_p$ lowered by a factor of $\chi$ we find that accretion
of small shear-dominated bodies from very thin disk is more 
important for embryo growth than accretion of large 
dispersion-dominated planetesimals whenever $\chi$ is such that 
condition (\ref{eq:acc_separ}) is fulfilled.


\section{Discussion.}
\label{sect:discus}


The outlined picture of planetesimal dynamics in the 
gaseous nebula naturally divides planetesimals of different 
sizes into two
well-defined populations with respect to gravitational scattering
by a set of protoplanetary cores. One is a ``hot''
population of bodies bigger than $0.1-1$ km which 
interact with embryos in the 
dispersion-dominated regime and have large inclinations
so that their scattering is an intrinsically  
three-dimensional process. 
The other is a ``cold'' population of smaller planetesimals
(sizes below 0.1-1 km) which interact with embryos in 
the shear-dominated regime. These planetesimals tend to 
be confined by the action of the gas drag 
to a vertically thin disk [with a thickness
of $\sim 10^{-3}R_H$] near the nebular 
midplane. 

Difference between these populations is most striking when 
the accretion of planetesimals by the embryos is concerned.
Accretion of dynamically hot planetesimals allows $10^{25}$ g embryo 
to double its mass in $\sim 10^5$ yr in the terrestrial 
planet region and in several $10^7$ yr in the region of 
ice giants. The latter timescale is quite long from 
the cosmogonical point of view. 
At the same time, accretion of cold 
planetesimals is about 100 
times faster, with mass doubling timescale of
$\sim 10^3$ yr and $\sim 10^5$ yr in the 
inner and outer parts of the nebula, 
provided that {\it all} solid mass is locked in these 
small planetesimals. Formation of gaseous atmospheres
around massive protoplanetary cores further accelerates 
accretion of small bodies: 
gaseous envelopes can be very efficient at 
trapping small planetesimals (Inaba \& Ikoma 2003), thus  
increasing the capture radius $R_e$ 
and shortening the growth timescale. Protoplanetary cores would also grow
by merging with other cores (since in our picture 
they reside on closely spaced orbits as long as 
$M_e\lesssim M_{e,crit}$, see 
\S \ref{subsect:migration}); apparently, the faster cores grow, the 
closer their orbits are in the Hill coordinates, and 
the faster they merge.   

It is possible that most of the solid mass in the proto-Solar 
nebula was concentrated from the very start in small 
($10-100$ m) dynamically cold bodies. In this case $\chi\approx 1$ 
and growth of embryos should be very fast, 
see equation (\ref{eq:t_grow_sd}).
It is, however, equally possible that most
of the mass was initially locked up in large ($1-10$ km) 
dispersion-dominated planetesimals and not in the dynamically cold 
population of small bodies, i.e. $\chi\ll 1$. 
In this case, although there is a huge reservoir of solid 
mass potentially available for the accretion, embryos can 
hardly make use of it because hot planetesimals are 
accreted rather inefficiently, see equation (\ref{eq:t_grow_dd}). 
At the same time, cold population which
can potentially allow a vigorous growth of the 
embryo might simply not contain enough surface density 
to ensure high enough accretion rate
(accretion timescale is inversely proportional to the 
fraction of mass contained in small planetesimals, 
see [\ref{eq:t_grow_sd}]). Thus, unless enough mass 
($\chi\gtrsim 10^{-2}$) is transferred from hot to 
cold planetesimals, embryos would grow slowly accreting 
large dispersion-dominated planetesimals.

A natural process for transferring 
mass from the big bodies into small ones is planetesimal
fragmentation (see Figure \ref{fig:mass_spectrum}), 
which should naturally be taking place in the presence 
of massive embryos. Indeed, the escape speed from the surface 
of 1 km body is about 1 m s$^{-1}$, while the Hill velocity of
$10^{25}$ g embryo is $\approx 50$ m s$^{-1}$ at 1 AU and 
 is $\approx 10$ m s$^{-1}$ at 30 AU. Thus, 1-km planetesimals
possibly  
containing most of the solid mass in the disk would collide 
with kinetic energy far exceeding their gravitational binding 
energy. Depending on their internal strength, parent 
bodies can be disrupted into a number of smaller fragments in  
such collisions. Collision strength 
is likely to be very small for objects in the outer Solar System 
which are thought to be composed primarily of ices. Comets 
are presumably 
the closest existing analogs of distant planetesimals 
and they are known to have small internal strength, e.g. from 
observations of tidal disruption of Shoemaker-Levy comet by  
Jupiter (Greenberg \etal 1995). Thus, it would be natural to
expect that collisional fragmentation triggered by the 
dynamical excitation of planetesimals by massive 
protoplanetary cores readily occurs at least in the outer 
Solar System. 

Efficiency of fragmentation is set by the collision timescale  
$t_{inel}$ of planetesimals with sizes in which most of 
the solid mass is concentrated. Using (\ref{eq:t_inel}) for 1 
km bodies with $\chi=1$ and $v_z\approx v_h$ we 
estimate\footnote{Collision timescale would be smaller for 
small $v_z/v_h$ (see \S \ref{sect:lower}), but this is 
partly compensated by the burst-like character of the 
fragmentation process, see below.} it to be 
$\sim 10^3$ yr at 1 AU and $\sim 10^7$ yr at 30 AU. 
This might seem like a rather long timescale in the outer 
Solar System but one should keep in mind that channeling just 
$10\%$ of mass into the population of cold planetesimals 
would increase the embryo's accretion rate by a factor of 
$\approx 10$ compared to the accretion of 
dispersion-dominated bodies, 
and this can be accomplished in a time 10 times shorter
than $t_{inel}$, i.e. in about several Myrs. Note 
that according to (\ref{eq:t_grow_sd}) the 
characteristic growth time of $10^{25}$ g embryo
by accretion of small bodies is also several Myrs 
for $\chi\approx 0.1$. Growth time decreases as $\chi$ 
gradually goes up meaning that
several Myrs is a natural evolution timescale 
for such embryos in the outer Solar System. 
We may conclude that if embryos grow mainly by 
accretion of small planetesimals then the planet 
formation timescale is intimately related to the 
timescale of fragmentation of massive planetesimals 
in catastrophic collisions. 

Accretion rate and dynamics of protoplanetary cores would in 
the end depend on the details of the time evolution 
of the mass fraction in small planetesimals $\chi$,
see \S \ref{subsect:migration} and \ref{sect:accretion}. 
Scaling of $\chi$ with time would also
determine whether inelastic collisions between small
planetesimals are an important dynamical 
factor; as we demonstrated in \S 
\ref{subsect:inelastic}, this can in some cases be an 
issue in the inner parts of the protoplanetary nebula. 
The amount of solid material contained in small 
planetesimals is 
determined by (1) the input of mass due to the 
fragmentation of large
planetesimals, (2) the removal of mass via the 
accretion by embryos,
and (3) the evolution of the surface density of small 
planetesimals due to the random scattering by embryos and 
their inward migration induced by the gas drag (Weidenschilling 1977;
Thommes \etal 2003). Self-consistent calculation 
of $\chi$ has to combine all these contributions and is 
beyond the scope of this study. 

Rapid accretion of small planetesimals 
can proceed provided that not only small bodies  
but also embryos themselves are on almost circular and 
uninclined orbits. As we have demonstrated in 
\S \ref{subsect:migration}, oligarchic growth allows 
simultaneous existence of many protoplanetary cores 
only if cores are lighter than about $10^{25}-10^{27}$ 
g (at 1 AU), since in that case their random motions can 
be kept small by planetesimal dynamical friction.
The exact value of the maximum core mass at which dynamical 
stability is still possible sensitively depends on the 
amount of mass concentrated in small shear-dominated bodies,
see \S \ref{subsect:migration}. After reaching this mass embryos would 
be dynamically excited ($e,i\gtrsim R_H/a$) and even if small 
planetesimals can still be kept confined to a cold,  
thin disk, embryo's accretion would 
proceed in the dispersion-dominated regime rather  slowly
(because relative 
embryo-planetesimal velocity is increased 
above $\Omega R_H$).

Discrete nature of the shear-dominated scattering of
small planetesimals by the embryos is very important for 
accurate calculation of processes characterized by the 
energy threshold such as the disruption
of small planetesimals in catastrophic collisions. 
Usual continuous approximation suitable in the 
dispersion-dominated regime would not work in such cases 
because it characterizes planetesimal velocity by its 
average value which is either above or below the 
threshold, meaning that corresponding process is either 
always on or always off. In reality, planetesimal velocity
changes continuously between the encounters with embryos
from very large values (about $\Omega R_H$) to very small 
ones. As a result, planetesimal velocity can be above 
the threshold for some time and during this period 
corresponding process would operate
(see Figure \ref{fig:inclination}). Later on velocity 
would drop below the threshold and process would switch 
off. This is qualitatively different from what one 
would obtain using continuous description for the 
planetesimal scattering
in the shear-dominated regime. For the fragmentation 
of small bodies this implies that
planetesimal destruction in catastrophic collisions 
proceeds in bursts right after each passage 
of the embryo, when relative planetesimal velocities are 
high, but later on, when velocities are damped by the gas drag, 
collisions might not be energetic enough to continue 
fragmenting planetesimals.


\section{Conclusions.}
\label{sect:concl}


We have explored the details of planetesimal dynamics near 
protoplanetary embryos in the presence of gas drag.
We showed that large ($\gtrsim 1$ km in size) 
planetesimals  are kept in the dispersion-dominated 
regime as a result of scattering by protoplanetary cores 
scattering, although their average 
velocities are reduced by the gas drag. 
Bodies smaller than roughly $0.1-1$ km
(depending on the location in the nebula) interact with 
protoplanetary cores in the shear-dominated regime. Owing to
the action of the gas drag these planetesimals settle
into a geometrically thin layer near the nebular midplane;  between
consecutive encounters with the embryos they
experience strong velocity damping which 
allows them to approach embryos every time with the relative 
velocity comparable to the Hill velocity --- minimum velocity which  
can be achieved in a differentially rotating disk.

For different locations in the proto-Solar nebula we 
have determined  which planetesimals are only weakly 
affected by the gas drag and are in the 
dispersion-dominated regime, and which planetesimals 
are so strongly coupled to the gas that their velocities are
below the Hill velocity of the protoplanetary cores. 
Dynamical peculiarities of the shear-dominated regime
in the presence of gas lead to a very 
high efficiency of accretion of small bodies by 
the embryos. If the surface mass density of 
small bodies is high enough ($\gtrsim 1\%$ of the 
total surface density in solids)
their accretion would dominate 
the embryo's growth rate (relative to the 
accretion of more massive, dispersion-dominated planetesimals). 

Large planetesimals ($\gtrsim 1$ km in size) likely 
containing most of the mass in solids have large random velocities 
and are not very efficiently accreted by embryos. 
However, they inelastically collide with 
each other at high velocities and fragment into smaller 
pieces contributing to the population of small 
bodies. Thus, the embryo's growth by accretion of small 
shear-dominated planetesimals 
can be regulated by the fragmentation of bigger, 
dispersion-dominated bodies. Planetesimal fragmentation 
would probably 
be easiest in the outer Solar System where colliding 
bodies are mostly composed of ices and are therefore 
internally weak and susceptible to easy destruction.
The natural timescale for the growth of $10^{25}$ g 
protoplanetary embryos by accretion of small 
planetesimals turns out to be around several Myrs 
at $30$ AU from the Sun. Cores of ice giants can be formed
in $\sim 10^7$ yr after large planetesimals lose all their 
mass in catastrophic collisions to small debris which
can be effectively accreted by massive embryos. 
This scenario would work only 
if the population of protoplanetary cores formed as the 
outcome of oligarchic growth can be kept on almost 
non-inclined and circular orbits. As we demonstrated in 
\S \ref{subsect:migration}, this is possible only if 
embryo masses do not exceed a specific limit dictated by 
the efficiency of planetesimal dynamical friction, which 
sensitively depends on the planetesimal mass fraction 
$\chi$ locked up in small planetesimals and velocity of
large dispersion-dominated planetesimals.

Future work should address the issues of the self-consistent
evolution of the mass fraction $\chi$ contained in small bodies; 
the role of inelastic collisions 
using improved fragmentation physics (can be an issue in
the inner Solar System); the final fate of massive embryos 
which cannot be 
kept dynamically cold by the planetesimal dynamical 
friction, and so on.

\acknowledgements

I am grateful to Jeremy Goodman for always stressing to me
the importance of gas for the planetesimal dynamics. I
have greatly benefited from numerous discussions with Peter 
Goldreich who has been working on similar problems. Useful 
comments by
Yoram Lithwick and Re'em Sari incited additional 
clarifications in \S \ref{subsect:migration}.
Author is a Frank and Peggy Taplin Member at the IAS;
he is also supported by the 
W. M. Keck Foundation and NSF grant PHY-0070928.


\begin{center}
\begin{deluxetable}{ l l l l l }
\tablewidth{0pc}
\tablecaption{Core accretion and dynamical friction regimes. 
\label{table2}}
\tablehead{
\colhead{Region\tablenotemark{a}}&
\colhead{Conditions}&
\colhead{Dyn. friction\tablenotemark{b}}&
\colhead{Accretion\tablenotemark{b}}&
\colhead{$M_{e,cr}/M_{iso}$\tablenotemark{c}}
}
\startdata
I & $\chi>(\Omega R_H/V)^4,~p^{1/2}(\Omega R_H/V)^2$ & 
SD\tablenotemark{b} & SD & $\chi^{3/2}$  \\
II & $p^{1/2}(\Omega R_H/V)^2<\chi<(\Omega R_H/V)^4$ & 
DD\tablenotemark{b} & SD & $(\Omega R_H/V)^6$  \\
III & $\chi<(\Omega R_H/V)^4,~p^{1/2}(\Omega R_H/V)^2$ & 
DD & DD & $1$  \\
IV & $(\Omega R_H/V)^4<\chi<p^{1/2}(\Omega R_H/V)^2$ & 
SD & DD & $\chi^{3/2}(V/\Omega R_H)^6$  \\
\enddata
\end{deluxetable}
\tablenotetext{a}{Roman numerals correspond to 
different regions in Figure \ref{fig:acc_df}.}
\tablenotetext{b}{Shows which planetesimal population ---
dispersion-dominated (DD) or shear-dominated (SD) --- 
dominates dynamical friction and accretion rate of embryos.}
\tablenotetext{c}{$M_{e,cr}$ and $M_{iso}$ are given by 
(\ref{eq:constr1}) and (\ref{eq:m_iso_cold}) in regions
I \& II and (\ref{eq:constr2}) and (\ref{eq:m_iso_hot})  
in regions III \& IV.}
\end{center}

\end{document}